\pdfoutput=1
\documentclass[twocolumn,showkeys,showpacs,preprintnumbers,prd,superscriptaddress,nofootinbib]{revtex4-2}
\usepackage{graphicx}
\usepackage{epsf}
\usepackage{bm}
\usepackage{amsmath}
\usepackage{amsfonts}
\usepackage{amssymb}
\usepackage{epstopdf}
\usepackage{natbib}
\usepackage{hyperref}
\usepackage{color}
\usepackage{verbatim}
\usepackage{multirow}
\usepackage{bm}
\usepackage{hyperref}
\usepackage{url}
\usepackage{orcidlink}
\usepackage{wrapfig}
\usepackage{nicefrac, xfrac}
\usepackage[normalem]{ulem}

\newcommand{\refeq}[1]{Eq.~(\ref{eq:#1})}     
\newcommand{\refeqs}[2]{Eqs.~(\ref{eq:#1})--(\ref{eq:#2})}     
     
\newcommand{\reffig}[1]{Fig.~\ref{fig:#1}} 
\newcommand{\reffigs}[2]{Fig.~\ref{fig:#1}--\ref{fig:#2}} 
     
\newcommand{\reftab}[1]{Table~\ref{tab:#1}}     
\newcommand{\refsec}[1]{Sec.~\ref{sec:#1}}
\newcommand{\refapp}[1]{App.~\ref{app:#1}}
\newcommand{\code}[1]{\texttt{#1}}

\makeatletter\let\expandableinput\@@input\makeatother

\definecolor{WildStrawberry}{HTML}{EE2967}

\begin{document}

\title{Exploring the Growth Index \texorpdfstring{$\gamma_L$}{}: Insights from Different CMB Dataset Combinations and Approaches}

\author{Enrico Specogna}
\email{especogna1@sheffield.ac.uk}
\author{Eleonora Di Valentino}
\email{e.divalentino@sheffield.ac.uk}
\affiliation{School of Mathematics and Statistics, University of Sheffield, Hounsfield Road, Sheffield S3 7RH, United Kingdom} 

\author{Jackson Levi Said}
\email{jackson.said@um.edu.mt}
\affiliation{Institute of Space Sciences and Astronomy, University of Malta}
\affiliation{Department of Physics, University of Malta}

\author{Nhat-Minh Nguyen\orcidlink{0000-0002-2542-7233}} \email{nguyenmn@umich.edu}
\affiliation{Leinweber Center for Theoretical Physics, University of Michigan, 
450 Church St, Ann Arbor, MI 48109-1040}
\affiliation{Department of Physics, College of Literature, Science and the Arts, University of Michigan, 450 Church St, Ann Arbor, MI 48109-1040}

\begin{abstract}
In this study we investigate the growth index $\gamma_L$, which characterizes the growth of linear matter perturbations, while analysing different cosmological datasets. We compare the approaches implemented by two different patches of the cosmological solver \texttt{CAMB}: \code{MGCAMB}~\cite{Zhao:2008bn,Pogosian:2010tj,Hojjati:2011ix,Zucca:2019xhg,Wang:2023tjj} and \code{CAMB\_GammaPrime\_Growth}~\cite{Nguyen:2023fip,Wen:2023bcj}. In our analysis we uncover a deviation of the growth index from its expected $\Lambda$CDM value of $0.55$ when utilizing the Planck dataset, both in the \code{MGCAMB} case and in the \code{CAMB\_GammaPrime\_Growth} case, but in opposite directions. This deviation is accompanied by a change in the direction of correlations with derived cosmological parameters. However, the incorporation of CMB lensing data helps reconcile $\gamma_L$ with its $\Lambda$CDM value in both cases. Conversely, the alternative ground-based telescopes ACT and SPT consistently yield growth index values in agreement with $\gamma_L$$=0.55$. We conclude that the presence of the A$_{\mathrm{lens}}$ problem in the Planck dataset contributes to the observed deviations, underscoring the importance of additional datasets in resolving these discrepancies.
\end{abstract}



\maketitle

\section{Introduction}
\label{sec:introduction}

The standard model of cosmology embodies the most universally accepted concordance model for both astrophysical and cosmological regimes within an isotropic and homogeneous Universe~\cite{Peebles:2002gy,Mukhanov:991646}. In this model, cold dark matter (CDM) acts as a stabilizing agent for galaxies and their clusters~\cite{Carr:2016drx,LUX:2016ggv,Gaitskell:2004gd}, while the late-time accelerating expansion of the Universe is sourced by a cosmological constant $\Lambda$~\cite{Riess:1998cb,Perlmutter:1998np}. In this scenario, initial cosmic inflation drove the early Universe towards flatness that observes the cosmological principle~\cite{Guth:1980zm,Linde:1981mu}.

However, challenges persist within this framework. The cosmological constant continues to pose theoretical difficulties~\cite{Weinberg:1988cp,Copeland:2006wr}, while the direct detection of CDM remains problematic~\cite{LUX:2016ggv,Gaitskell:2004gd}, and the UV completeness of the theory remains an open question~\cite{Addazi:2021xuf}. In recent years, additional challenges to $\Lambda$CDM cosmology have emerged in the form of cosmic tensions.  
These tensions arise from discrepancies between direct measurements of the cosmic expansion rate and the growth of large scale structure (LSS), and those inferred indirectly from early Universe measurements~\cite{Abdalla:2022yfr,DiValentino:2020vhf,DiValentino:2020vvd,Staicova:2021ajb,DiValentino:2021izs,Perivolaropoulos:2021jda,DiValentino:2022oon,DiValentino:2020zio,SajjadAthar:2021prg,Nunes:2021ipq}.

The presence of cosmic tensions has become evident across multiple surveys, primarily manifested in the measurement of the Hubble constant, $H_0$, which has now exceeded $5\sigma$~\cite{Verde:2019ivm,DiValentino:2020vnx,Riess:2019qba,Riess:2022mme}. The indirect measurements of $H_0$ mainly rely on observations of the Cosmic Microwave Background radiation (CMB) with the latest reported value by the Planck collaboration giving $H_0 = 67.4 \pm 0.5 \,{\rm km\, s}^{-1} {\rm Mpc}^{-1}$~\cite{Planck:2018vyg}, while the last release from ACT puts $H_0 = 67.9 \pm 1.5 \,{\rm km\, s}^{-1} {\rm Mpc}^{-1}$~\cite{ACT:2020gnv}. Conversely, late time probes offer several alternatives for the direct measurement of the Hubble expansion of the Universe. Among these, the most precise measurements are based on Type Ia Supernovae (SNIa) by the SH0ES Team, calibrated by Cepheids, yielding $H_0 = 73.04 \pm 1.04 \,{\rm km\, s}^{-1} {\rm Mpc}^{-1}$~\cite{Riess:2021jrx}. Numerous other measurements also consistently favor higher values for the Hubble constant~\cite{Wong:2019kwg,Huang:2019yhh,Pesce:2020xfe,Kourkchi:2020iyz,Schombert:2020pxm,Blakeslee:2021rqi,deJaeger:2022lit,Shajib:2023uig}. 
Finally, the latest measurements employing the Tip of the Red Giant Branch as a calibration method for SNIa continue to indicate elevated values for $H_0$~\cite{Scolnic:2023mrv,Anderson:2023aga,Freedman:2021ahq,Freedman:2020dne}.

Another cosmological tension that is gaining significance relates to the growth of large-scale structures in the Universe. Several parameters capture these characteristics, with $S_8 = \sigma_{8,0} \sqrt{\Omega_{\rm m,0} / 0.3}$ being a key quantity that incorporates the matter density parameter at present $\Omega_{\rm m,0}$ and is commonly used to quantify it, as it is directly measured by Weak Lensing (WL) experiments. The measurement of $S_8$ is more challenging, with discrepancies between early- and late-time probes ranging from $2-3.5\sigma$, but it still plays a crucial role in testing the validity of $\Lambda$CDM cosmology. Recent measurements from the Kilo-Degree Survey (KiDS-1000) report a lower value of $S_8 = 0.766^{+0.020}_{-0.014}$~\cite{Heymans:2020gsg}, which is consistent with other WL measurements such as DES-Y3~\cite{DES:2021wwk} and HSC-Y3~\cite{Dalal:2023olq}. On the other hand, Planck (TT,TE,EE+lowE) gives $S_8 = 0.834 \pm 0.016$~\cite{Planck:2018vyg} which agrees with other CMB based measurements of this parameter~\cite{ACT:2020gnv,SPT-3G:2022hvq}.

The presence of cosmic tensions has spurred numerous re-evaluations in the scientific community, leading to more rigorous investigations of potential systematic errors~\cite{Brout:2021mpj,DiValentino:2020zio,Perivolaropoulos:2021jda}. However, the persistence of these tensions across various direct and indirect measurements suggests that such explanations may be losing credibility. Consequently, alternative modifications to the standard model of cosmology have been proposed, encompassing a range of theoretical frameworks (see for example~\cite{Knox:2019rjx,Jedamzik:2020zmd,DiValentino:2021izs,Perivolaropoulos:2021jda,Abdalla:2022yfr,Poulin:2023lkg,DiValentino:2022fjm,Addazi:2021xuf,Krishnan:2020vaf,Adil:2023jtu,CANTATA:2021ktz,Kamionkowski:2022pkx} and the references therein). This plethora of models that go beyond $\Lambda$CDM physics can be parametrized in numerous ways. By examining the phenomenological parameterization for the growth of linear matter perturbations $\delta_m = \delta\rho_m/\rho_m$ described in~\cite{Linder:2005in}, the growth rate $G(a)$ can be approximately reformulated as $G(a)=f(a, \gamma_L, \Omega_{m,0})$, a time-dependent function $f$ where the growth index $\gamma_L$ can be seen as a fitting constant~\cite{Linder:2003dr,Linder:2002et} (discussed further in Sec.~\ref{sec:theory}). This empirical formula gives a value of $\gamma_L\simeq0.55$ for $\Lambda$CDM~\cite{Linder:2005in,Wang:1998gt}.

This theoretical prediction for $\gamma_L$ allows us to use this parameter as a testing tool for possible deviations from the concordance model of cosmology. In this work, we analyze this parameterization through the prism of the linear matter perturbation evolution equation (Eqs. \ref{eq:delta_per_evo}-\ref{eq:growthG}) defined in Sec. \ref{sec:theory}. In Sec.~\ref{sec:codes} we include a discussion on the main differences in the implementation of $\gamma_L$ in the two cosmological solvers considered in our study: \texttt{MGCAMB} and \texttt{CAMB\_GammaPrime\_Growth}. In Sec.~\ref{sec:data} we discuss the data sets under investigation, along with the strategy that we followed to obtain the constraints on the $\Lambda\mathrm{CDM}+\gamma_L$ model considered here, which are outlined in Sec.~\ref{sec:res}, where the outcomes of both approaches are compared. In particular, the results for \texttt{MGCAMB}'s scale-dependent approach is shown in Sec.~\ref{sec:scale_de}, along with \texttt{CAMB\_GammaPrime\_Growth}'s scale-independent one in Sec.~\ref{sec:scale_inde}. Finally, the main results are summarized and discussed in Sec.~\ref{sec:conclusions}.

\section{Structure Formation and the Growth Index}
\label{sec:theory}

The dynamics of the expansion of the Universe can be determined through the Friedmann equation:
\begin{equation}
    \label{eq:friedmann}
    \left(\frac{H}{H_0}\right)^2=\sum_i \Omega_i\,,
\end{equation}
where $\Omega_i=\rho_i/\rho_c$ represents the relative density of each component with energy density $\rho_i$ in a flat Universe where $\rho_c=3H_0^2/(8\pi G_{GR})$, with $G_{GR}$ being Newton's constant of gravitation in General Relativity (GR).
Indeed we can use Eq.(\ref{eq:friedmann}) to constrain the observed accelerated expansion, but it has to reflect our ignorance of the fundamental nature of such acceleration. Since we do not know whether we are dealing with a physical energy density or rather a modification of the Friedmann equation with respect to the $\Lambda$CDM model, we can treat the accelerated expansion as stemming from an effective component with equation of state $w(a)$, rewriting Eq.(\ref{eq:friedmann}) as follows~\cite{Linder:2003dr,Linder:2004ng}:
\begin{align}
\label{eq:friedmann_mod}
    \left(\frac{H}{H_0}\right)^2 = \Omega_{\rm m,0}a^{-3} + (1 - \Omega_{\rm m,0}) e^{3\int_0^{a'} \left[1+w(a')\right]d\ln a'}\,.
\end{align}
However, background expansion alone cannot easily tell different cosmological theories apart, as functions like $w(a)$ can be tuned in different cosmologies to obtain the same time evolution for expansion history quantities such as $a(t)$~\cite{Linder:2005in}. For this reason, in this paper we will assume a $\Lambda$CDM background; namely, $w(a) = -1$ in Eq.~\ref{eq:friedmann_mod}.\\ \\

Instead, we consider modifications to the growth of structure \textit{only}. The evolution of the expansion rate feeds into the inhomogeneities that arise in the LSS due to early Universe gravitational instabilities and frictional terms. By investigating perturbations away from homogeneity and isotropy in both the gravitational and matter sectors, we can describe the matter perturbation evolution equation as
\begin{equation}\label{eq:delta_per_evo}
    \ddot{\delta}_m + 2H \dot{\delta}_m - 4\pi \rho_m \delta_m = 0\,,
\end{equation}
where $\delta_m$ represents the gauge invariant fractional matter density, dots denote derivatives with respect to cosmic time, and where linear and subhorizon scales ($k \lesssim 0.1 h {\rm Mpc}^{-1}$ and $k\gtrsim0.0003h{\rm Mpc}^{-1}$ respectively, $h=H_0/100\,{\rm kms}^{-1}{\rm Mpc}^{-1}$) are being considered. A natural definition of the linear growth fraction for matter perturbations is $g(a) = \delta_m(a)/\delta_{m,0}$, where $\delta_{m,0}$ is the current value of $\delta_m$. By taking a reparameterization of the matter perturbation evolution equation through the growth rate $G(a) = d\ln g(a)/d\ln a$, Eq.~\eqref{eq:delta_per_evo} can be rewritten as~\cite{Linder:2018pth}
\begin{equation}
\label{eq:growthG}
    \frac{dG}{d\ln a} + G^2 + \left(2 + \frac{1}{2}\frac{d\ln H^2}{d\ln a}\right) G - \frac{3}{2} G_{\rm eff}\Omega_{\rm m} (a) = 0\,,
\end{equation}
where $G_{\rm eff}$ is $G_{\rm MG}/G_{\rm GR}$, and $G_{\rm MG}$ is Newton's constant of gravitation for a generic deviation from GR.
It was proposed in~\cite{Linder:2007hg} that
\begin{equation}
\label{eq:gamma_param}
    G(a) = \Omega_{\rm m}^{\gamma_L} (a)\,,
\end{equation}
where $\gamma_L = \ln G/\ln \Omega_{\rm m} (a)$ is the \textit{growth index}. A major advantage of this parameterization is its accuracy in replicating the numerical solution of eq.(\ref{eq:growthG}). For instance, in the framework of $\Lambda$CDM, the value $\gamma_L^{\rm \Lambda CDM} = 0.55$ can be as accurate as $0.05\%$~\cite{Linder:2005in}.\\

An interesting approach considered in the literature has been to take the background Friedmann equations to be those of $\Lambda$CDM and to probe potential modifications away from the standard model at the level of $\gamma_L$~\cite{Linder:2003ze,Linder:2023klx,Linder:2006xb}. In fact, for theories other than GR such as $f(R)$ and DGP braneworld cosmologies we get remarkably different values: $\gamma_L^{f(R)} = 0.42$ and $\gamma_L^{\rm DGP} = 0.68$ respectively~\cite{Linder:2023klx}. A departure of $\gamma_L$ from $0.55$ would then represent a possible hint at new physics, offering us an opportunity to shed new light on the tensions afflicting the standard model of cosmology.

\section{Mapping the growth index into observables: \texttt{MGCAMB} and $\texttt{CAMB\_GammaPrime\_Growth}$}
\label{sec:codes}

To constrain the growth index $\gamma_L$, we first need a mean to map it into the CMB and LSS observables, where the latter are specifically probed by CMB lensing potentials in this work. Here, we adopt two different modifications of the standard Boltzmann solver \code{CAMB}~\cite{Lewis:1999bs,Howlett:2012mh}, namely \texttt{MGCAMB}~\cite{Zhao:2008bn,Pogosian:2010tj,Hojjati:2011ix,Zucca:2019xhg,Wang:2023tjj} and \texttt{CAMB\_GammaPrime\_Growth}~\cite{Nguyen:2023fip,Wen:2023bcj}. Below, we briefly review the main differences between the two approaches.

In the first approach, \code{MGCAMB} introduces the two scale- and redshift-dependent parameters $\mu(a,k)$ and $\eta(a,k)$, which fully describe the relations between the $\Psi$- $\Phi$ metric perturbations and the energy-momentum tensor of GR, as can be observed from Eqs.~(5)-(9) in~\cite{Pogosian:2010tj} (with $\eta=\mu=1$ corresponding to GR and $\Lambda$CDM).
Separating sub- and super-horizon evolutions of perturbations, \code{MGCAMB} then solves the two regimes separately~\cite{Pogosian:2010tj}.
\code{MGCAMB} maps $\gamma_L$ into the function $\mu(a,k)$ through the following relation, valid for subhorizon perturbations~\cite{Pogosian:2010tj,Hojjati:2011ix}:

\begin{equation}
    \label{eq:MGCAMB_gamma_mu}
    \mu=\frac{2}{3}\Omega_{\rm m}^{\gamma_L-1}\left[\Omega_{\rm m}^{\gamma_L}+2+\frac{H'}{H}+\gamma_L\frac{\Omega'_{\rm m}}{\Omega_{\rm m}}+\gamma_L'\ln(\Omega_{\rm m})\right].
\end{equation}
Therefore, within \code{MGCAMB}, any change in $\gamma_L$ is effectively a change in $\mu(a,k)$, where the latter appears in the evolution of both sub- and super-horizon perturbations, as described by Eqs.~(16)-(17) and Eq.~(21) of~\cite{Pogosian:2010tj}.
This choice implies that varying $\gamma_L$ introduces a \emph{scale-dependent} effect on observables such as the primary CMB angular power spectrum and the linear matter power spectrum.
Ref.~\cite{Pogosian:2010tj} demonstrated that, for a simple toy model where $\mu(a,k)=\mu(a)$, super-horizon changes in the linear matter power spectrum introduced by a varying $\mu(a)$ are within the cosmic variance. They thereby argued that \refeq{MGCAMB_gamma_mu} and their approach to $\gamma_L$ should be consistent with the (implicit) assumptions of scale-independence and subhorizon evolution by \refeq{gamma_param}.
In practice, we however find that this is not necessarily the case for the (primary) CMB angular power spectrum. In \refapp{code_comparison}, we further show why this particular choice might explain the constraint on $\gamma_L$ obtained by \code{MGCAMB} in Sect.~\ref{sec:scale_de}.\footnote{See also Ref.~\cite{Xu:2013tsa,Sakr:2023bms} for previous analyses with Planck data.}

The second code, \code{CAMB\_GammaPrime\_Growth}~\cite{Nguyen:2023fip,gamma_prime_growth,Wen:2023bcj}, adopts a phenomenological approach by rescaling the linear matter power spectrum $P(k)$, i.e.
\begin{equation}
\label{eq:CAMB_GammaPrime_Growth_gamma_Pk}
P(\gamma_L,k,a)=P(k,a=1)\, D^2(\gamma_L,a),
\end{equation}
by the (square of) the modified growth factor $D(\gamma_L,a)$ given by the numerical integral
\begin{equation}
\label{eq:CAMB_GammaPrime_Growth_gamma_Da}
D(\gamma_L,a)=\exp\left[-\int_a^1\,da\,\frac{\Omega_{\rm m}^{\gamma_L}(a)}{a}\right].
\end{equation}
For the CMB observables and data considered in this work, \refeqs{CAMB_GammaPrime_Growth_gamma_Pk}{CAMB_GammaPrime_Growth_gamma_Da} only affect the CMB lensing potential through the matter transfer function (see, e.g. Eqs.~(9)-(14) in~\cite{Hanson:2009kr}). We will return to this point in \refapp{code_comparison} where we compare this approach by \code{CAMB\_GammaPrime\_Growth} versus that by \code{MGCAMB}, and their respective constraints on $\gamma_L$ using Planck(+lensing) data.


\section{Datasets and Methodology}
\label{sec:data}


Together with the growth index $\gamma_L$ defined in Sec.\ref{sec:theory}, we constrained the six parameters of the base $\Lambda$CDM model. The parameter space constrained here is therefore $\Gamma\equiv\{\Omega_\mathrm{b} h^2, \Omega_\mathrm{c} h^2, 100\theta_\mathrm{MC}, \tau_\mathrm{reio}, n_\mathrm{s}, \log(10^{10} A_\mathrm{s}), \gamma_L\} $, where $\Omega_bh^2$, the baryon density $\Omega_b$ combined with the dimensionless Hubble constant $h$, $\Omega_ch^2$, the density of cold dark matter combined with $h$, $100\theta_{MC}$, where $\theta_{MC}$ is an approximation of the observed angular size of the sound horizon at recombination, $\tau_\mathrm{reio}$, the reionization optical depth, $n_s$ and $\log(10^{10} A_\mathrm{s})$, respectively the spectral index and amplitude $A_\mathrm{s}$ of the power spectrum of the scalar primordial perturbations.\\

We constrained the $\Lambda$CDM$+\gamma_L$ model considered here with two different methods:
\begin{itemize}
    \item Firstly, we used \texttt{MGCosmoMC}~\cite{Zhao:2008bn,Pogosian:2010tj,Hojjati:2011ix,Zucca:2019xhg,Wang:2023tjj}, a publicly available modification of the popular MCMC code \texttt{CosmoMC}~\cite{Lewis:2002ah,Lewis:2013hha} that samples $\gamma_L$ through the fast-slow dragging algorithm developed in~\cite{neal2005taking}. The Einstein-Boltzmann solver implemented in \texttt{MGCosmoMC} to calculate CMB anisotropies and lensing power spectra is \texttt{MGCAMB}~\cite{Zhao:2008bn,Pogosian:2010tj,Hojjati:2011ix,Zucca:2019xhg,Wang:2023tjj}, a patch of the \texttt{CAMB} code~\cite{Lewis:1999bs} that includes several phenomenological modifications to the growth of matter perturbations, including the $\gamma_L$ parameterization in Eq.(\ref{eq:gamma_param}).
    \item Additionally, we implemented the modified version of \texttt{CAMB}, namely \code{CAMB\_GammaPrime\_Growth}, introduced in~\cite{Nguyen:2023fip,gamma_prime_growth,Wen:2023bcj} into \texttt{CosmoMC}. While~\cite{Nguyen:2023fip,gamma_prime_growth,Wen:2023bcj} also modify linear growth through the parameterization shown in Eq.\ref{eq:gamma_param}, they take $\gamma_L$ to be scale-independent compared to \texttt{MGCAMB}.
\end{itemize}

The data considered in our analysis have been taken from the following datasets:
\begin{itemize}
    \item the final release of the Planck mission~\cite{Planck:2018vyg,Planck:2018nkj,Planck:2019nip}, specifically the likelihoods and data for the CMB temperature and polarization anisotropies: TT, EE and TE (referred to, here, as \textit{Planck}), along with the lensing reconstruction power spectrum (referred to as \textit{lensing});
    \item  the likelihoods and data from the  9-years release (DR9) of the Wilkinson Microwave Anisotropy Probe (\textit{WMAP})~\cite{WMAP:2012fli};
    \item the likelihoods for the polarization and temperature anisotropies spectra from Atacama Cosmology Telescope's DR4 (\textit{ACT})~\cite{ACT:2020frw};
    \item the EE and TE anisotropies measurements and likelihoods from the SPT-3G instrument of the South Pole Telescope (\textit{SPT})~\cite{SPT-3G:2021eoc};
    \item the final measurements from the SDSS survey by the eBOSS collaboration, of which we include Baryonic Acoustic Oscillations (\textit{BAO}) data~\cite{eBOSS:2020yzd}.
\end{itemize}

In particular, we constrained $\Gamma$ with the following combinations: 
 \begin{itemize}
     \item Planck, Planck$+$BAO, Planck$+$lensing, and Planck$+$BAO$+$lensing;
     \item ACT, ACT$+$WMAP, ACT$+$BAO and ACT$+$WMAP+BAO;
     \item SPT, SPT$+$WMAP, SPT$+$BAO and SPT$+$WMAP$+$BAO.
 \end{itemize}
 
We note that all the constraints obtained using either the ACT or SPT data employed a Gaussian prior on $\tau_{reio}=0.065\pm0.015$, as done by the ACT collaboration~\cite{ACT:2020frw}; the other six parameters making up the $\Lambda$CDM$+\gamma_L$ model considered here were all assumed to have flat priors, as summarised in Table \ref{tab:priors} .\\

\begin{table}
\resizebox{0.42\columnwidth}{!}{
\begin{tabular}{c | c}
\hline
\textbf{Parameter} & \textbf{Prior} \\ 
\hline\hline

$ \Omega_\mathrm{b} h^2  $ & $[0.005, 0.1]$
\\ 
$ \Omega_\mathrm{c} h^2  $  
& $[0.001, 0.99]$
\\ 
$ 100\theta_\mathrm{MC}  $  & $[0.5, 10]$
\\ 

$ \tau_\mathrm{reio}  $  & $0.065\pm0.015$
\\ 
$ n_\mathrm{s}  $  & $[0.8, 1.2]$
\\ 
$ \log(10^{10} A_\mathrm{s})  $  & $[1.61, 3.91]$
 \\ 
$ \gamma_L  $ & $[0, 1]$
 \\

\hline \hline
\end{tabular} }
\caption{A summary of the priors imposed on the parameters for the $\Lambda$CDM$+\gamma_L$ model considered in our analysis.}
\label{tab:priors}
\end{table}

The convergence of the chains produced from the exploration of $\Gamma$ by \texttt{MGCosmoMC/CosmoMC} for this analysis was evaluated through the Gelman-Rubin criterion~\cite{gelman}, and we took $R\leq 0.02$ as a satisfactory limit at which to present the results outlined in Sec.\ref{sec:res}.\\

Finally, to demonstrate that a Planck-like CMB only experiment is able to recover $\gamma_L = 0.55$ without bias, we constrained $\Gamma$ using a mock Planck-like dataset generated according to a $\Lambda$CDM model defined by the parameter choices given in Table \ref{tab:plfake}. The results are shown in Table \ref{tab:mock_res}, and we can see that CMB only data are able to recover $\gamma_L = 0.55$ with good accuracy with either \texttt{MGCAMB} or \texttt{CAMB\_GammaPrime\_Growth}.\\

\begin{table}[h!]
\resizebox{0.5\columnwidth}{!}{
    \begin{tabular}{c|c}
        \hline
        \textbf{Parameter} & \textbf{Fiducial Values} \\ 
        \hline\hline
        
        $ \Omega_\mathrm{b} h^2  $ & $0.02236$
        \\ 
        $ \Omega_\mathrm{c} h^2  $  
        & $0.1202$
        \\ 
        $ 100\theta_\mathrm{MC}  $  & $1.04090$
        \\ 
        
        $ \tau_\mathrm{reio}  $  & $0.0544$
        \\ 
        $ n_\mathrm{s}  $  & $0.9649$
        \\ 
        $ \log(10^{10} A_\mathrm{s})  $  & $3.044$
         \\
        
        \hline \hline
    \end{tabular}}
    \caption{The $\Lambda$CDM baseline values chosen to simulate mock Planck-like dataset used for the results of Table \ref{tab:mock_res}.}
    \label{tab:plfake}
\end{table}

\begin{table}[h!]
\resizebox{0.9\columnwidth}{!}{
    \begin{tabular}{c|c|c}

\hline
\textbf{Parameter} & \textbf{\texttt{MGCAMB}} & \textbf{\texttt{CAMB\_GammaPrime\_Growth}}\\
\hline \hline
$\Omega_b h^2$ &  $ 0.02237\pm 0.00015 $&$ 0.02236\pm 0.00016$
\\

$\Omega_c h^2$ &  $ 0.1201\pm 0.0016 $&$ 0.1202\pm 0.0017$
\\

$100\theta_{MC} $ &  $ 1.04091\pm 0.00036 $&$ 1.04089\pm 0.00036$
\\

$\tau$ &  $ 0.0538\pm 0.0097 $&$ 0.053\pm 0.010$
\\
$n_s$ &  $ 0.9650\pm 0.0041 $&$ 0.9649\pm 0.0043$
\\
$\log(10^{10} A_\mathrm{s})$ &  $ 3.044\pm 0.019 $&$ 3.043\pm 0.021$
\\
$\gamma_L$ &  $ 0.544^{+0.035}_{-0.021} $&$ 0.56\pm 0.10$ \\
\hline \hline
    \end{tabular}}
    \caption{Results for the Planck-like dataset.}
    \label{tab:mock_res}
\end{table}


\section{Results}
\label{sec:res}

In this section we are discussing the constraints we obtained for the dataset combinations listed in the previous \autoref{sec:data} and the two different approaches explained in \autoref{sec:codes}. In particular we will discuss in \autoref{sec:scale_de} the scale-dependent case as implemented in \code{MGCAMB}, and in \autoref{sec:scale_inde} the scale-independent linear growth as implemented in $\texttt{CAMB\_GammaPrime\_Growth}$.


\subsection{MGCAMB Results}
\label{sec:scale_de}


\begin{table*}
\begin{center}
\renewcommand{\arraystretch}{1.5}
\resizebox{0.7\textwidth}{!}{
\begin{tabular}{l c c c c c c c c c c c c c c c }
\hline
\textbf{Parameter} & \textbf{ Planck} & \textbf{ Planck+BAO}& \textbf{ Planck+lensing} & \textbf{ Planck+BAO+lensing} \\ 
\hline\hline

$ \Omega_\mathrm{b} h^2  $ & $ 0.02253\pm 0.00016 $&$ 0.02250\pm 0.00016 $&$ 0.02249\pm 0.00016 $&$ 0.02246\pm 0.00015 $

  \\ 
$ \Omega_\mathrm{c} h^2  $ & $ 0.1187\pm 0.0015 $&$ 0.1190\pm 0.0013 $&$ 0.1186\pm 0.0014 $&$ 0.1189\pm 0.0012 $

 \\ 
$ 100\theta_\mathrm{MC}  $ & $ 1.04110\pm 0.00032 $&$ 1.04106\pm 0.00032 $&$ 1.04107\pm 0.00032 $&$ 1.04103\pm 0.00030 $

  \\ 
$ \tau_{reio}  $& $ 0.0510\pm 0.0085 $&$ 0.0507\pm 0.0082 $&$ 0.0496^{+0.0087}_{-0.0073} $&$ 0.0490^{+0.0083}_{-0.0073} $

\\ 

$ n_\mathrm{s}  $  & $ 0.9688\pm 0.0047 $&$ 0.9681\pm 0.0045 $&$ 0.9684\pm 0.0046 $&$ 0.9675\pm 0.0042 $

  \\ 
$ \log(10^{10} A_\mathrm{s})  $  & $ 3.034\pm 0.018 $&$ 3.034\pm 0.017 $&$ 3.030^{+0.018}_{-0.015} $&$ 3.030\pm 0.017 $

 \\ 
$ \gamma_L  $ & $ 0.467^{+0.018}_{-0.029} $&$ 0.469^{+0.017}_{-0.029} $&$ 0.506\pm 0.022 $&$ 0.509^{+0.022}_{-0.020} $

 \\
\hline
$ H_0  $  & $ 68.02\pm 0.66 $&$ 67.86\pm 0.60 $&$ 68.00\pm 0.64 $&$ 67.84\pm 0.57 $

 \\ 
$S_8$  & $ 0.839\pm 0.015 $&$ 0.842\pm 0.015 $&$ 0.824\pm 0.013 $&$ 0.827\pm 0.012 $

 \\

\hline \hline
\end{tabular} }
\end{center}
\caption{Constraints at 68\% CL for the scale-dependent case as implemented in \code{MGCAMB}, considering the dataset combinations with Planck.}
\label{tab:Pl-MG}
\end{table*}



\begin{table*}
\begin{center}
\renewcommand{\arraystretch}{1.5}
\resizebox{0.7\textwidth}{!}{
\begin{tabular}{l c c c c c }
\hline
\textbf{Parameter} & \textbf{ ACT} & \textbf{ ACT+BAO}& \textbf{ ACT+WMAP}& \textbf{ ACT+WMAP+BAO}  \\ 
\hline\hline

$ \Omega_\mathrm{b} h^2  $  & $0.02153\pm 0.00032 $&$ 0.02148\pm 0.00030 $&$ 0.02239\pm 0.00021 $&$ 0.02237\pm 0.00019$

  \\ 
$ \Omega_\mathrm{c} h^2  $  & $0.1177\pm 0.0049 $&$ 0.1200\pm 0.0025 $&$ 0.1198\pm 0.0027 $&$ 0.1203\pm 0.0020$

 \\ 
$ 100\theta_\mathrm{MC}  $  & $1.04229\pm 0.00078 $&$ 1.04210\pm 0.00065 $&$ 1.04174\pm 0.00066 $&$ 1.04171\pm 0.00062$

  \\ 
$ \tau_{reio}  $  & $0.064\pm 0.015 $&$ 0.065\pm 0.015 $&$ 0.060\pm 0.012 $&$ 0.060\pm 0.012$

  \\ 
$ n_\mathrm{s}  $   & $1.008\pm 0.017 $&$ 1.003\pm 0.014 $&$ 0.9729\pm 0.0062 $&$ 0.9722\pm 0.0052$

  \\ 
$ \log(10^{10} A_\mathrm{s})  $   & $3.049\pm 0.034 $&$ 3.056\pm 0.033 $&$ 3.062^{+0.025}_{-0.022} $&$ 3.064^{+0.025}_{-0.022}$

 \\ 
$ \gamma_L  $  & $0.552^{+0.063}_{-0.087} $&$ 0.580^{+0.062}_{-0.078} $&$ 0.533^{+0.044}_{-0.018} $&$ 0.536^{+0.040}_{-0.018}$

 \\
\hline
$ H_0  $   & $68.0\pm 2.0 $&$ 67.0\pm 1.0 $&$ 67.7\pm 1.1 $&$ 67.50\pm 0.83$

 \\ 
$S_8$   & $0.829\pm 0.045 $&$ 0.847\pm 0.028 $&$ 0.843\pm 0.031 $&$ 0.848\pm 0.025$

 \\

\hline \hline
\end{tabular} }
\end{center}
\caption{Constraints at 68\% CL for the scale-dependent case as implemented in \code{MGCAMB}, considering the dataset combinations with ACT.}
\label{tab:ACT-MG }
\end{table*}



\begin{table*}
\begin{center}
\renewcommand{\arraystretch}{1.5}
\resizebox{0.7\textwidth}{!}{
\begin{tabular}{l c c c c c }
\hline
\textbf{Parameter} & \textbf{ SPT} & \textbf{ SPT+BAO}& \textbf{ SPT+WMAP}& \textbf{ SPT+WMAP+BAO}  \\ 
\hline\hline

$ \Omega_\mathrm{b} h^2  $  & $0.02238\pm 0.00033 $&$ 0.02237\pm 0.00032 $&$ 0.02264\pm 0.00023 $&$ 0.02259\pm 0.00021$

  \\ 
$ \Omega_\mathrm{c} h^2  $  & $0.1175\pm 0.0057 $&$ 0.1186\pm 0.0026 $&$ 0.1153\pm 0.0028 $&$ 0.1171\pm 0.0020$

 \\ 
$ 100\theta_\mathrm{MC}  $  & $1.03945\pm 0.00081 $&$ 1.03933\pm 0.00069 $&$ 1.03973\pm 0.00066 $&$ 1.03954\pm 0.00064$

  \\ 
$ \tau_{reio}  $ & $0.065\pm 0.015 $&$ 0.066\pm 0.015 $&$ 0.060\pm 0.013 $&$ 0.058\pm 0.013$

  \\ 
$ n_\mathrm{s}  $   & $0.991\pm 0.025 $&$ 0.987\pm 0.019 $&$ 0.9733\pm 0.0075 $&$ 0.9709\pm 0.0067$

  \\ 
$ \log(10^{10} A_\mathrm{s})  $   & $3.040\pm 0.039 $&$ 3.043\pm 0.038 $&$ 3.041\pm 0.025 $&$ 3.042\pm 0.026$

 \\ 
$ \gamma_L  $  & $0.622^{+0.075}_{-0.11} $&$ 0.635^{+0.063}_{-0.084} $&$ 0.556^{+0.023}_{-0.018} $&$ 0.558^{+0.024}_{-0.018}$

 \\
\hline
$ H_0  $   & $67.8\pm 2.3 $&$ 67.3\pm 1.0 $&$ 68.9\pm 1.2 $&$ 68.11\pm 0.83$

 \\ 
$S_8$   & $0.796\pm 0.048 $&$ 0.804\pm 0.028 $&$ 0.782\pm 0.032 $&$ 0.801\pm 0.025$

 \\

\hline \hline
\end{tabular} }
\end{center}
\caption{Constraints at 68\% CL for the scale-dependent case as implemented in \code{MGCAMB}, considering the dataset combinations with SPT.}
\label{tab:SPT-MG }
\end{table*}



\begin{figure*}
   \includegraphics[width=0.7\textwidth]{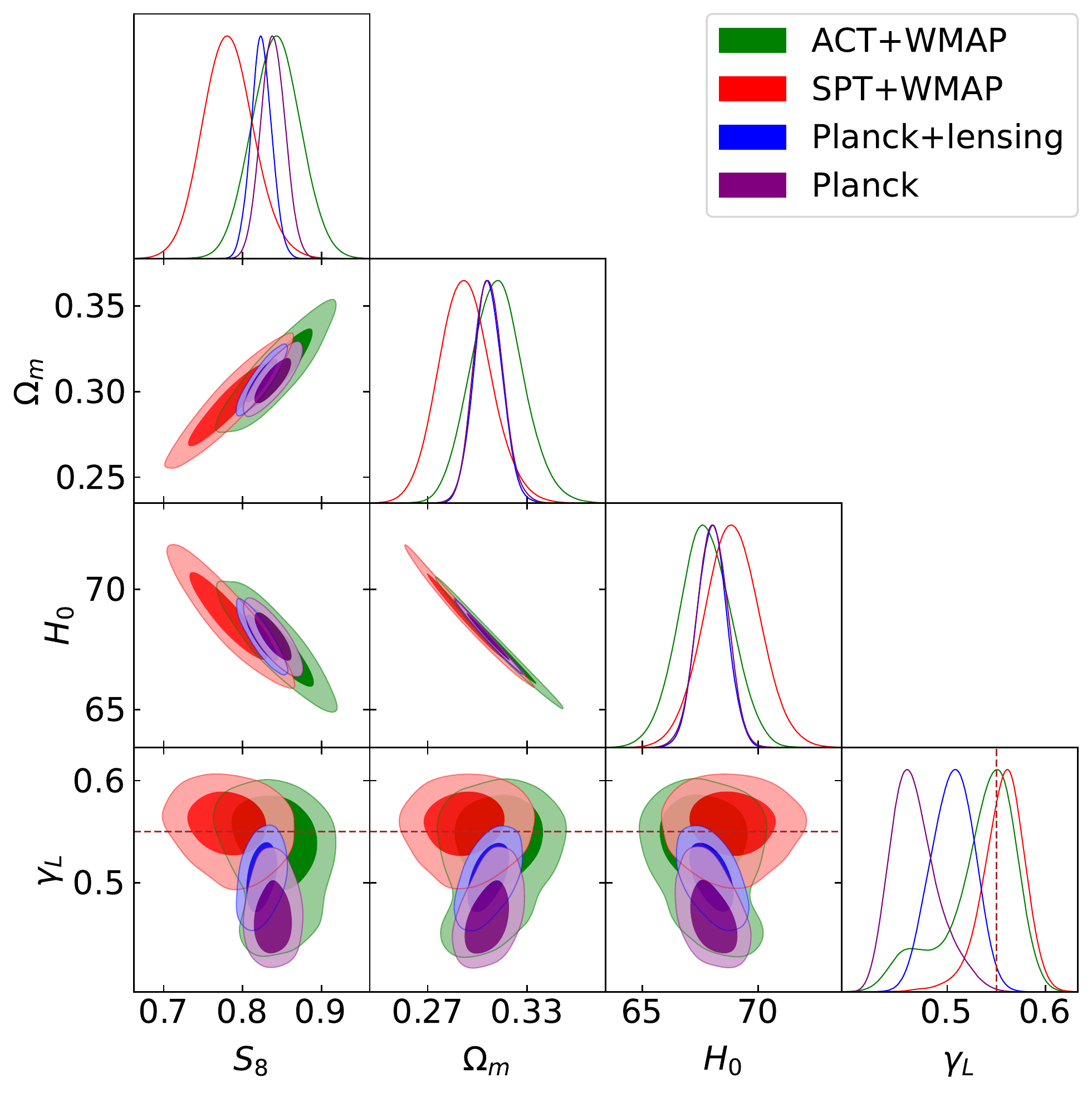}
     \caption{1D marginalized posterior distributions and 2D contour plots for the different CMB data combinations explored in this work, without the inclusion of the BAO data, for the scale-dependent case as implemented in \code{MGCAMB}. The dashed brown line in the plot is the value of the growth index in $\Lambda$CDM, $\gamma_L=0.55$.}
    \label{fig:MG-noBAO}
\end{figure*}



\begin{figure*}
   \includegraphics[width=0.7\textwidth]{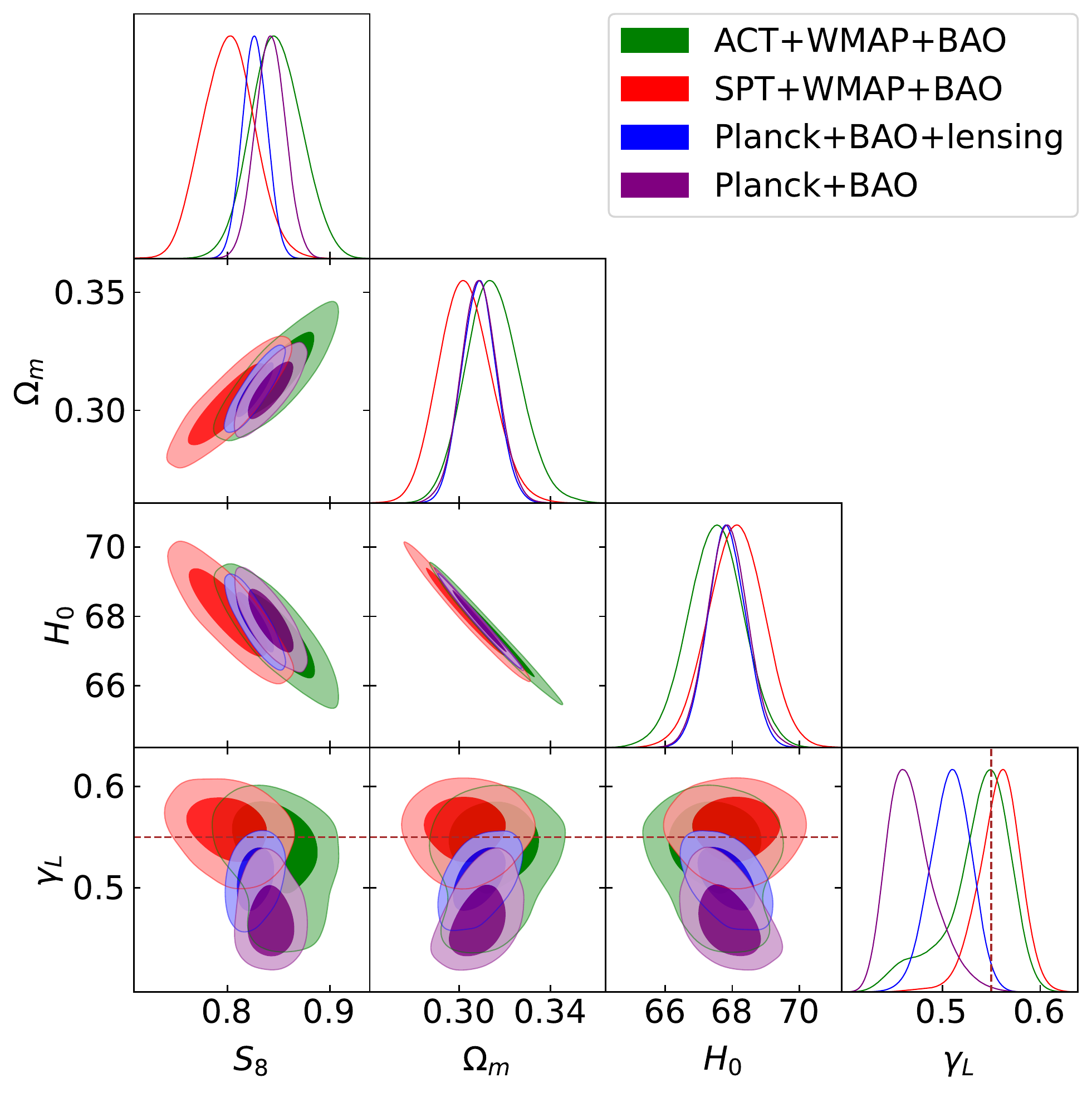}
     \caption{1D marginalized posterior distributions and 2D contour plots for the different CMB+BAO data combinations explored in this work, for the scale-dependent case as implemented in \code{MGCAMB}. The dashed brown line in the plot is the value of the growth index in $\Lambda$CDM, $\gamma_L=0.55$.}
    \label{fig:MG-BAO}
\end{figure*}


We show in \autoref{tab:Pl-MG} the constraints at 68\% CL for the scale-dependent case as implemented in \code{MGCAMB} for the different combinations involving the Planck data, and we repeat the same for ACT in \autoref{tab:ACT-MG } and SPT in \autoref{tab:SPT-MG }. We display instead the 1D posterior distributions and the 2D contour plots for all these cases in \autoref{fig:MG-noBAO} and \autoref{fig:MG-BAO}.

The analysis presented in \autoref{tab:Pl-MG} reveals a notable deviation from the expected value of $\gamma_L=0.55$ considering the measurement obtained solely from Planck observations. Specifically, Planck indicates a value of $\gamma_L=0.467^{+0.018}_{-0.029}$, deviating from the expected value by more than 3 standard deviations. However, it is important to note that this parameter does not contribute significantly to resolving the cosmological tensions. While it exhibits a slight correlation with $H_0$, resulting in a modest increase of only $1\sigma$ in its value ($H_0=68.02\pm0.66$ km/s/Mpc), it does not exhibit any correlation with $S_8$ (see \autoref{fig:MG-noBAO}). Notably, the disagreement between WL measurements (which assume a $\Lambda$CDM model) and $S_8$ persists, further emphasizing the unresolved tensions in the current cosmological framework.
Even when incorporating BAO data, the same conclusion holds true. The inclusion of BAO data does not alter the constraints on $\gamma_L$ and $S_8$, but it does lead to a slight decrease in the estimated value of $H_0$. However, a more significant impact is observed when including the lensing dataset. Specifically, with Planck+lensing the constraint on $\gamma_L$ becomes $\gamma_L=0.506\pm0.022$, reducing the tension with the expected value to $2\sigma$. Furthermore, a mild correlation emerges between the $\gamma_L$ parameter and the $S_8$ parameter (see \autoref{fig:MG-noBAO}). This correlation causes the mean value of $S_8$ to shift downward by 1 standard deviation, albeit not sufficiently enough to alleviate the existing $S_8$ tension. Once again, the inclusion of BAO data does not change the results, so that Planck+BAO+lensing gives similar constraints to the Planck+lensing combination.

A different picture emerges when the ground based CMB telescope ACT is taken into account in \autoref{tab:ACT-MG }. In this case, the value of $\gamma_L$ remains consistently within 1 standard deviation of the expected value across all dataset combinations, including also BAO and WMAP data. Furthermore, by referring to \autoref{fig:MG-noBAO} and \autoref{fig:MG-BAO}, we observe a change: the correlation between $\gamma_L$ and other cosmological parameters disappears upon inclusion of the ACT data.

Similarly, when the SPT data are analyzed in \autoref{tab:SPT-MG } we observe that across all combinations of datasets $\gamma_L$ is always in agreement within $1\sigma$ with $\gamma_L=0.55$. In particular, when considering SPT data alone, or in combination with BAO data, higher values of $\gamma_L$ are favored with larger error bars. However, upon including the WMAP dataset, the mean value of $\gamma_L$ returns to $\gamma_L=0.55$.  Similar to the previous analysis, no correlation is observed between $\gamma_L$ and the derived parameters depicted in \autoref{fig:MG-noBAO} and \autoref{fig:MG-BAO} when examining this dataset.


\subsection{\code{CAMB\_GammaPrime\_Growth} Results}
\label{sec:scale_inde}


\begin{table*}
\begin{center}
\renewcommand{\arraystretch}{1.5}
\resizebox{0.7\textwidth}{!}{
\begin{tabular}{l c c c c c c c c c c c c c c c }
\hline
\textbf{Parameter} & \textbf{ Planck} & \textbf{ Planck+BAO}& \textbf{ Planck+lensing} & \textbf{ Planck+BAO+lensing} \\ 
\hline\hline

$ \Omega_\mathrm{b} h^2  $  & $0.02258\pm 0.00016 $&$ 0.02255\pm 0.00016 $&$ 0.02251\pm 0.00017 $&$ 0.02248\pm 0.00016 $
  \\ 
$ \Omega_\mathrm{c} h^2  $  & $0.1181\pm 0.0015 $&$ 0.1186\pm 0.0013 $&$ 0.1183\pm 0.0015 $&$ 0.1188\pm 0.0013 $
 \\ 
$ 100\theta_\mathrm{MC}  $ & $1.04113\pm 0.00032 $&$ 1.04108\pm 0.00031 $&$ 1.04109\pm 0.00032 $&$ 1.04103\pm 0.00032 $\\ 

$ \tau_\mathrm{reio}  $   & $0.0496^{+0.0087}_{-0.0074} $&$ 0.0495\pm 0.0084 $&$ 0.0493^{+0.0087}_{-0.0074} $&$ 0.0488^{+0.0086}_{-0.0075} $
\\ 
$ n_\mathrm{s}  $  & $0.9709\pm 0.0047 $&$ 0.9696\pm 0.0045 $&$ 0.9696\pm 0.0048 $&$ 0.9683\pm 0.0044 $
  \\ 
$ \log(10^{10} A_\mathrm{s})  $  & $3.030\pm 0.017 $&$ 3.031\pm 0.018 $&$ 3.029^{+0.018}_{-0.016} $&$ 3.029^{+0.018}_{-0.016} $
 \\ 
$ \gamma_L  $ & $0.841^{+0.11}_{-0.074} $&$ 0.831^{+0.11}_{-0.080} $&$ 0.669\pm 0.069 $&$ 0.658\pm 0.063 $
 \\
\hline
$ H_0  $  & $68.27\pm 0.69 $&$ 68.06\pm 0.61 $&$ 68.14\pm 0.70 $&$ 67.92\pm 0.61 $
 \\ 
$S_8$  & $0.805\pm 0.018 $&$ 0.810\pm 0.017 $&$ 0.807\pm 0.019 $&$ 0.812\pm 0.017 $
 \\

\hline \hline
\end{tabular} }
\end{center}
\caption{Constraints at 68\% CL for the \code{CAMB\_GammaPrime\_Growth} case as implemented in~\cite{Nguyen:2023fip,gamma_prime_growth,Wen:2023bcj}, considering the dataset combinations with Planck.}
\label{tab:Pl-Minh}
\end{table*}



\begin{table*}
\begin{center}
\renewcommand{\arraystretch}{1.5}
\resizebox{0.7\textwidth}{!}{
\begin{tabular}{l c c c c c }
\hline
\textbf{Parameter} & \textbf{ ACT} & \textbf{ ACT+BAO}& \textbf{ ACT+WMAP}& \textbf{ ACT+WMAP+BAO}  \\ 
\hline\hline

$ \Omega_\mathrm{b} h^2  $  & $0.02151\pm 0.00031 $&$ 0.02149\pm 0.00031 $&$ 0.02236\pm 0.00021 $&$ 0.02235\pm 0.00020$

  \\ 
$ \Omega_\mathrm{c} h^2  $  & $0.1185\pm 0.0050 $&$ 0.1199\pm 0.0026 $&$ 0.1205\pm 0.0030 $&$ 0.1206\pm 0.0022$

 \\ 
$ 100\theta_\mathrm{MC}  $  & $1.04221\pm 0.00077 $&$ 1.04205\pm 0.00065 $&$ 1.04168\pm 0.00066 $&$ 1.04164\pm 0.00060$

  \\ 
$ \tau_{reio}  $  & $0.065\pm 0.015 $&$ 0.065\pm 0.015 $&$ 0.062\pm 0.013 $&$ 0.061\pm 0.013$

  \\ 
$ n_\mathrm{s}  $   & $1.007\pm 0.017 $&$ 1.003\pm 0.014 $&$ 0.9722\pm 0.0066 $&$ 0.9720\pm 0.0055$

  \\ 
$ \log(10^{10} A_\mathrm{s})  $   & $3.053\pm 0.034 $&$ 3.057\pm 0.032 $&$ 3.067\pm 0.026 $&$ 3.067\pm 0.026$

 \\ 
$ \gamma_L  $  & $0.52\pm 0.18 $&$ 0.49\pm 0.15 $&$ 0.50\pm 0.16 $&$ 0.50\pm 0.14$

 \\
\hline
$ H_0  $   & $67.7\pm 2.0 $&$ 67.1\pm 1.0 $&$ 67.4\pm 1.3 $&$ 67.39\pm 0.89$

 \\ 
$S_8$   & $0.839\pm 0.059 $&$ 0.855\pm 0.032 $&$ 0.847\pm 0.036 $&$ 0.848\pm 0.027$

 \\

\hline \hline
\end{tabular} }
\end{center}
\caption{Constraints at 68\% CL for the \code{CAMB\_GammaPrime\_Growth} case as implemented in~\cite{Nguyen:2023fip,gamma_prime_growth,Wen:2023bcj}, considering the dataset combinations with ACT.}
\label{tab:ACT-Minh}
\end{table*}



\begin{table*}
\begin{center}
\renewcommand{\arraystretch}{1.5}
\resizebox{0.7\textwidth}{!}{
\begin{tabular}{l c c c c c }
\hline
\textbf{Parameter} & \textbf{ SPT} & \textbf{ SPT+BAO}& \textbf{ SPT+WMAP}& \textbf{ SPT+WMAP+BAO}  \\ 
\hline\hline

$ \Omega_\mathrm{b} h^2  $  & $0.02241\pm 0.00033 $&$ 0.02238\pm 0.00031 $&$ 0.02259\pm 0.00024 $&$ 0.02253\pm 0.00022$

  \\ 
$ \Omega_\mathrm{c} h^2  $  & $0.1164\pm 0.0056 $&$ 0.1183\pm 0.0026 $&$ 0.1167\pm 0.0032 $&$ 0.1178\pm 0.0021$

 \\ 
$ 100\theta_\mathrm{MC}  $  & $1.03953\pm 0.00081 $&$ 1.03935\pm 0.00067 $&$ 1.03955\pm 0.00071 $&$ 1.03942\pm 0.00064$

  \\ 
$ \tau_{reio}  $   & $0.065\pm 0.015 $&$ 0.065\pm 0.015 $&$ 0.062\pm 0.013 $&$ 0.061\pm 0.013$

  \\ 
$ n_\mathrm{s} $   & $0.994\pm 0.024 $&$ 0.989\pm 0.019 $&$ 0.9709\pm 0.0080 $&$ 0.9687\pm 0.0068$

 \\ 

 $\log(10^{10} A_\mathrm{s})$  & $3.035\pm 0.039 $&$ 3.040\pm 0.035 $&$ 3.049\pm 0.027 $&$ 3.051\pm 0.027$

 \\
 
 $\gamma_L  $  & $0.46\pm 0.19 $&$ 0.41\pm 0.15 $&$ 0.43\pm 0.14 $&$ 0.41\pm 0.13$

 \\
\hline
$ H_0  $   & $68.3^{+2.1}_{-2.4} $&$ 67.4\pm 1.0 $&$ 68.3\pm 1.4 $&$ 67.77\pm 0.89$

 \\ 
$S_8$   & $0.803\pm 0.064 $&$ 0.824\pm 0.032 $&$ 0.802\pm 0.039 $&$ 0.815\pm 0.027$

 \\

\hline \hline
\end{tabular} }
\end{center}
\caption{Constraints at 68\% CL for the \code{CAMB\_GammaPrime\_Growth} case as implemented in~\cite{Nguyen:2023fip,gamma_prime_growth,Wen:2023bcj}, considering the dataset combinations with SPT.}
\label{tab:SPT-Minh}
\end{table*}



\begin{figure*}[t]
   \includegraphics[width=0.7\textwidth]{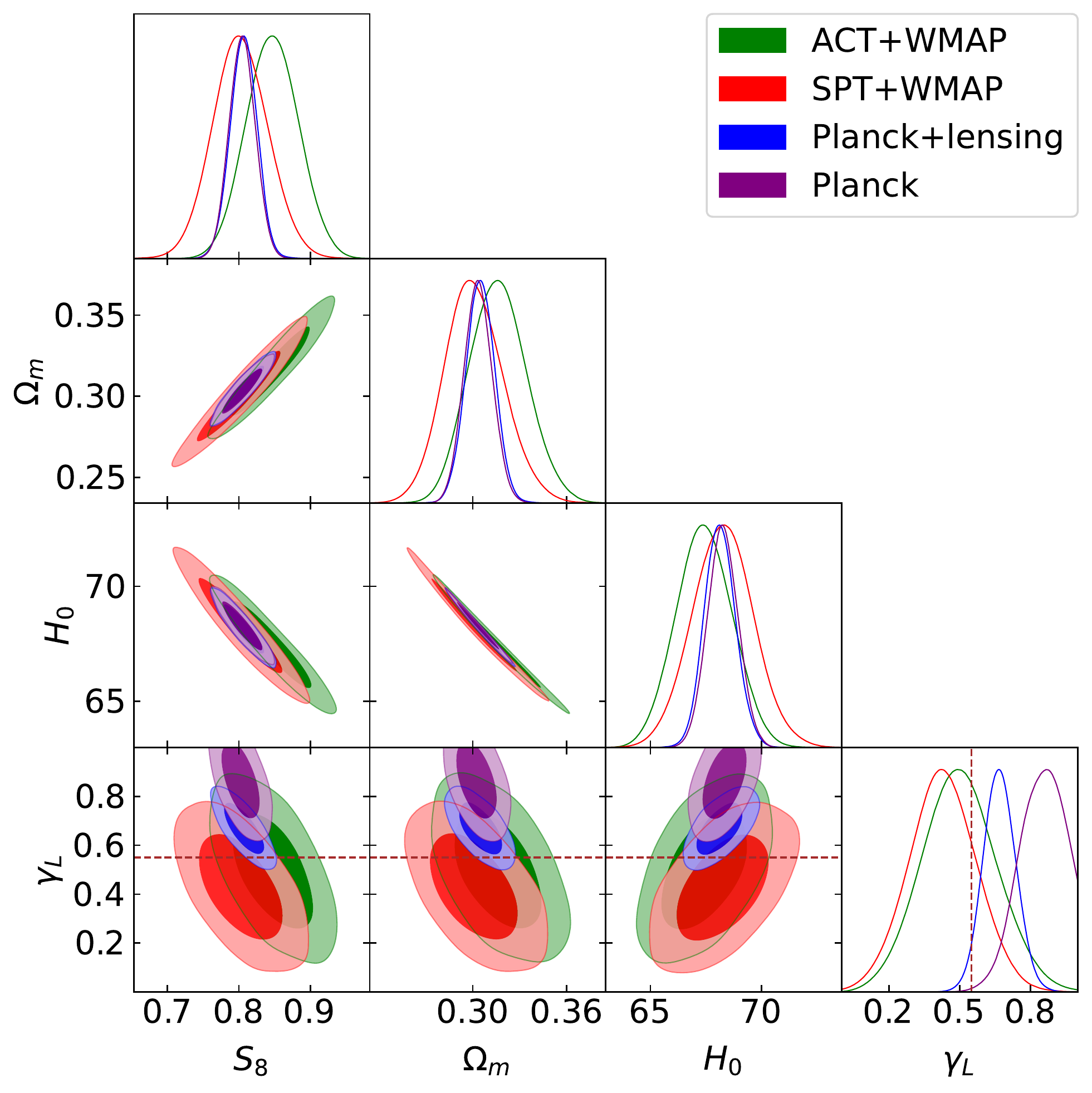}
     \caption{1D marginalized posterior distributions and 2D contour plots for the different CMB data combinations explored in this work, without the inclusion of the BAO data, for the \code{CAMB\_GammaPrime\_Growth} case as implemented in~\cite{Nguyen:2023fip,gamma_prime_growth,Wen:2023bcj}. The dashed brown line in the plot is the value of the growth index in $\Lambda$CDM, $\gamma_L=0.55$.}
    \label{fig:Minh-noBAO}
\end{figure*}



\begin{figure*}[t]
   \includegraphics[width=0.7\textwidth]{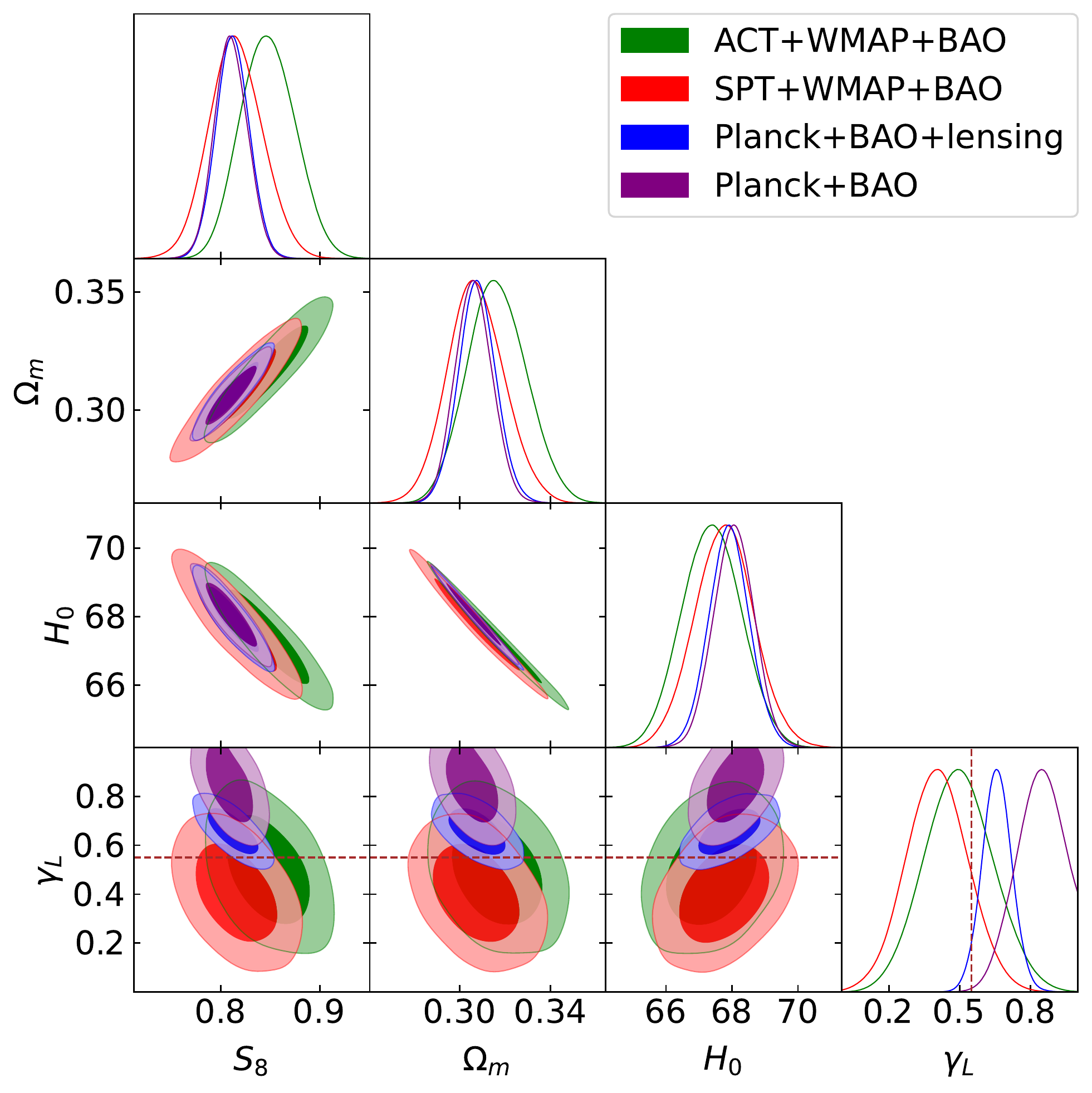}
     \caption{1D marginalized posterior distributions and 2D contour plots for the different CMB+BAO data combinations explored in this work, for the \code{CAMB\_GammaPrime\_Growth} case as implemented in~\cite{Nguyen:2023fip,gamma_prime_growth,Wen:2023bcj}. The dashed brown line in the plot is the value of the growth index in $\Lambda$CDM, $\gamma_L=0.55$.}
    \label{fig:Minh-BAO}
\end{figure*}


The constraints at a confidence level of 68\% CL for the \code{CAMB\_GammaPrime\_Growth} case as implemented in~\cite{Nguyen:2023fip,gamma_prime_growth,Wen:2023bcj},
are presented in \autoref{tab:Pl-Minh}, for the various combinations involving Planck data. Similarly, we provide the corresponding constraints for ACT in \autoref{tab:ACT-Minh} and for SPT in \autoref{tab:SPT-Minh}. To further illustrate these cases, we present the 1D posterior distributions and the 2D contour plots in \autoref{fig:Minh-noBAO} and \autoref{fig:Minh-BAO}.

The examination of \autoref{tab:Pl-Minh} reveals a significant deviation from the expected value of $\gamma_L=0.55$ when considering the measurement based solely on Planck observations. Specifically, Planck indicates a value of $\gamma_L=0.841^{+0.11}_{-0.074}$, surpassing the expected value by more than 3$\sigma$. However, it is worth noting that while this parameter does not contribute significantly to resolving the $H_0$ tension, it can substantially decrease the value of the $S_8$ parameter in the right direction to agree with the WL measurements (which assume a $\Lambda$CDM model). In fact, contrarily to the \code{MGCAMB} case, $\gamma_L$ exhibits a slight correlation with both $H_0$ and $S_8$ (see \autoref{fig:Minh-noBAO}), giving $H_0=68.27\pm0.69$ km/s/Mpc and $S_8=0.805\pm0.018$. It is important to emphasize that in the \code{CAMB\_GammaPrime\_Growth} case, there are two notable distinctions. Firstly, the preferred value of $\gamma_L$ deviates from the expected value in the opposite direction compared to the \code{MGCAMB} case. Secondly, the correlations between $\gamma_L$, $H_0$, and $S_8$ exhibit a change in sign.
However, as we noted in the previous section as well, the inclusion of the CMB lensing dataset is crucial to shift the value of $\gamma_L$ towards the expected $\gamma_L=0.55$, reducing the disagreement at the level of $1.7\sigma$. In particular we obtain $\gamma_L=0.669\pm0.069$, in perfect agreement with~\cite{Nguyen:2023fip}, while leaving both $H_0$ and $S_8$ unaffected. Finally, the addition of the BAO data does not have a significant impact on either the Planck+BAO combination or the Planck+BAO+lensing combination.

If we now consider the independent CMB measurements obtained from ACT displayed in \autoref{tab:ACT-Minh} we observe a similar pattern as in the previous section. Notably, $\gamma_L$ returns to agree within $1\sigma$ with the expected value $\gamma_L=0.55$. In this case the addition of BAO and WMAP data does not alter these conclusions. In this particular dataset combination, $\gamma_L$ exhibits a negative correlation with $S_8$ and a positive degeneracy with $H_0$. However, their mean values remain robust and align with the values expected in a $\Lambda$CDM model, in absence of deviations in $\gamma_L$ from $0.55$. The same conclusions about constraints and parameter degeneracies remain valid when replacing ACT with SPT, as demonstrated in \autoref{tab:SPT-Minh}.

In conclusion, the observed deviation of $\gamma_L$ when analyzing the Planck data can be attributed to the presence of the $A_{\mathrm{lens}}$ problem~\cite{Planck:2018vyg,Calabrese:2008rt,DiValentino:2020hov}, as highlighted in~\cite{Nguyen:2023fip}. This problem refers to the excess of lensing detected in the temperature power spectrum, which is also associated with indications of a closed Universe~\cite{Planck:2018vyg,DiValentino:2019qzk,Handley:2019tkm}. The mitigation of the $\gamma_L$ deviation from $0.55$ upon incorporating the CMB lensing data serves as direct evidence supporting this interpretation.


\section{Conclusions}
\label{sec:conclusions}

We investigated the growth index $\gamma_L$, which characterizes the growth of linear matter perturbations in the form shown in \refeq{gamma_param}, by using different cosmological datasets, and comparing two approaches for the implementation of $\gamma_L$ into \texttt{CAMB}: the \code{MGCAMB} and \code{CAMB\_GammaPrime\_Growth} codes.\\

Our analysis in the \code{MGCAMB} case revealed a $\gamma_L$ that deviates from its $\Lambda$CDM value of $0.55$, preferring instead lower values and indicating a discrepancy of more than 3 standard deviations. However, incorporating the CMB lensing dataset helps reduce this disagreement to approximately 2 standard deviations. 

In the \code{CAMB\_GammaPrime\_Growth} case instead, the preferred value of $\gamma_L$ differs from the \code{MGCAMB} case, and exceeds the expected value in the opposite direction, producing a change of sign of the correlations with $H_0$ and $S_8$. Similarly to \texttt{MGCAMB}, for the \code{CAMB\_GammaPrime\_Growth} case the CMB lensing dataset helps in reconciling $\gamma_L$ with the expected $0.55$ value. 

Moreover, the analysis of the ACT and SPT datasets shows consistent agreement of $\gamma_L$ with the expected value within 1 standard deviation across various dataset combinations, and the addition of BAO data has minimal impact on the constraints and parameter correlations in both the \code{MGCAMB} and \code{CAMB\_GammaPrime\_Growth} cases. 

Given these facts, we can attribute the deviation of $\gamma_L$ observed in the Planck dataset to the $A_{\mathrm{lens}}$ problem (as already noticed in~\cite{Nguyen:2023fip}) characterized by excess lensing in the temperature power spectrum.

Overall, these findings highlight the importance of considering additional datasets, such as CMB lensing, and other experiments, such as ACT and SPT, when tackling apparent discrepancies with the standard model, such as the deviations of $\gamma_L$ encountered in this work.

\begin{acknowledgments}
\noindent  

This article is based upon work from COST Action CA21136 Addressing observational tensions in cosmology with systematics and fundamental physics (CosmoVerse) supported by COST (European Cooperation in Science and Technology).
We would like to thank the referee for their constructive feedback and helpful comments.
We acknowledge IT Services at The University of Sheffield for the provision of services for High Performance Computing. The \code{MGCAMB}-\code{CAMB\_GammaPrime\_Growth} code verification and comparison was performed on the Great Lakes HPC cluster, maintained by the Advanced Research Computing division, UofM Information and Technology Service. We further thank Dragan Huterer, Levon Pogosian, Alessandra Silvestri and Zhuangfei (Xavier) Wang for helpful discussions related to \code{MGCAMB} and their implementation of $\gamma_L$. EDV is supported by a Royal Society Dorothy Hodgkin Research Fellowship. MN acknowledges support from the Leinweber Center for Theoretical Physics, the NASA grant under contract 19-ATP19-0058, and the DOE under contract DE-FG02-95ER40899. JLS would also like to acknowledge funding from ``The Malta Council for Science and Technology'' as part of the REP-2023-019 (CosmoLearn) Project.
\end{acknowledgments}

\bibliography{references}

\begin{thebibliography}{92}%
\makeatletter
\providecommand \@ifxundefined [1]{%
 \@ifx{#1\undefined}
}%
\providecommand \@ifnum [1]{%
 \ifnum #1\expandafter \@firstoftwo
 \else \expandafter \@secondoftwo
 \fi
}%
\providecommand \@ifx [1]{%
 \ifx #1\expandafter \@firstoftwo
 \else \expandafter \@secondoftwo
 \fi
}%
\providecommand \natexlab [1]{#1}%
\providecommand \enquote  [1]{``#1''}%
\providecommand \bibnamefont  [1]{#1}%
\providecommand \bibfnamefont [1]{#1}%
\providecommand \citenamefont [1]{#1}%
\providecommand \href@noop [0]{\@secondoftwo}%
\providecommand \href [0]{\begingroup \@sanitize@url \@href}%
\providecommand \@href[1]{\@@startlink{#1}\@@href}%
\providecommand \@@href[1]{\endgroup#1\@@endlink}%
\providecommand \@sanitize@url [0]{\catcode `\\12\catcode `\$12\catcode
  `\&12\catcode `\#12\catcode `\^12\catcode `\_12\catcode `\%12\relax}%
\providecommand \@@startlink[1]{}%
\providecommand \@@endlink[0]{}%
\providecommand \url  [0]{\begingroup\@sanitize@url \@url }%
\providecommand \@url [1]{\endgroup\@href {#1}{\urlprefix }}%
\providecommand \urlprefix  [0]{URL }%
\providecommand \Eprint [0]{\href }%
\providecommand \doibase [0]{https://doi.org/}%
\providecommand \selectlanguage [0]{\@gobble}%
\providecommand \bibinfo  [0]{\@secondoftwo}%
\providecommand \bibfield  [0]{\@secondoftwo}%
\providecommand \translation [1]{[#1]}%
\providecommand \BibitemOpen [0]{}%
\providecommand \bibitemStop [0]{}%
\providecommand \bibitemNoStop [0]{.\EOS\space}%
\providecommand \EOS [0]{\spacefactor3000\relax}%
\providecommand \BibitemShut  [1]{\csname bibitem#1\endcsname}%
\let\auto@bib@innerbib\@empty
\bibitem [{\citenamefont {Zhao}\ \emph {et~al.}(2009)\citenamefont {Zhao},
  \citenamefont {Pogosian}, \citenamefont {Silvestri},\ and\ \citenamefont
  {Zylberberg}}]{Zhao:2008bn}%
  \BibitemOpen
  \bibfield  {author} {\bibinfo {author} {\bibfnamefont {G.-B.}\ \bibnamefont
  {Zhao}}, \bibinfo {author} {\bibfnamefont {L.}~\bibnamefont {Pogosian}},
  \bibinfo {author} {\bibfnamefont {A.}~\bibnamefont {Silvestri}},\ and\
  \bibinfo {author} {\bibfnamefont {J.}~\bibnamefont {Zylberberg}},\ }\href
  {https://doi.org/10.1103/PhysRevD.79.083513} {\bibfield  {journal} {\bibinfo
  {journal} {Phys. Rev. D}\ }\textbf {\bibinfo {volume} {79}},\ \bibinfo
  {pages} {083513} (\bibinfo {year} {2009})},\ \Eprint
  {https://arxiv.org/abs/0809.3791} {arXiv:0809.3791 [astro-ph]} \BibitemShut
  {NoStop}%
\bibitem [{\citenamefont {Pogosian}\ \emph {et~al.}(2010)\citenamefont
  {Pogosian}, \citenamefont {Silvestri}, \citenamefont {Koyama},\ and\
  \citenamefont {Zhao}}]{Pogosian:2010tj}%
  \BibitemOpen
  \bibfield  {author} {\bibinfo {author} {\bibfnamefont {L.}~\bibnamefont
  {Pogosian}}, \bibinfo {author} {\bibfnamefont {A.}~\bibnamefont {Silvestri}},
  \bibinfo {author} {\bibfnamefont {K.}~\bibnamefont {Koyama}},\ and\ \bibinfo
  {author} {\bibfnamefont {G.-B.}\ \bibnamefont {Zhao}},\ }\href
  {https://doi.org/10.1103/PhysRevD.81.104023} {\bibfield  {journal} {\bibinfo
  {journal} {Phys. Rev. D}\ }\textbf {\bibinfo {volume} {81}},\ \bibinfo
  {pages} {104023} (\bibinfo {year} {2010})},\ \Eprint
  {https://arxiv.org/abs/1002.2382} {arXiv:1002.2382 [astro-ph.CO]}
  \BibitemShut {NoStop}%
\bibitem [{\citenamefont {Hojjati}\ \emph {et~al.}(2011)\citenamefont
  {Hojjati}, \citenamefont {Pogosian},\ and\ \citenamefont
  {Zhao}}]{Hojjati:2011ix}%
  \BibitemOpen
  \bibfield  {author} {\bibinfo {author} {\bibfnamefont {A.}~\bibnamefont
  {Hojjati}}, \bibinfo {author} {\bibfnamefont {L.}~\bibnamefont {Pogosian}},\
  and\ \bibinfo {author} {\bibfnamefont {G.-B.}\ \bibnamefont {Zhao}},\ }\href
  {https://doi.org/10.1088/1475-7516/2011/08/005} {\bibfield  {journal}
  {\bibinfo  {journal} {JCAP}\ }\textbf {\bibinfo {volume} {08}},\ \bibinfo
  {pages} {005}},\ \Eprint {https://arxiv.org/abs/1106.4543} {arXiv:1106.4543
  [astro-ph.CO]} \BibitemShut {NoStop}%
\bibitem [{\citenamefont {Zucca}\ \emph {et~al.}(2019)\citenamefont {Zucca},
  \citenamefont {Pogosian}, \citenamefont {Silvestri},\ and\ \citenamefont
  {Zhao}}]{Zucca:2019xhg}%
  \BibitemOpen
  \bibfield  {author} {\bibinfo {author} {\bibfnamefont {A.}~\bibnamefont
  {Zucca}}, \bibinfo {author} {\bibfnamefont {L.}~\bibnamefont {Pogosian}},
  \bibinfo {author} {\bibfnamefont {A.}~\bibnamefont {Silvestri}},\ and\
  \bibinfo {author} {\bibfnamefont {G.-B.}\ \bibnamefont {Zhao}},\ }\href
  {https://doi.org/10.1088/1475-7516/2019/05/001} {\bibfield  {journal}
  {\bibinfo  {journal} {JCAP}\ }\textbf {\bibinfo {volume} {05}},\ \bibinfo
  {pages} {001}},\ \Eprint {https://arxiv.org/abs/1901.05956} {arXiv:1901.05956
  [astro-ph.CO]} \BibitemShut {NoStop}%
\bibitem [{\citenamefont {Wang}\ \emph {et~al.}(2023)\citenamefont {Wang},
  \citenamefont {Mirpoorian}, \citenamefont {Pogosian}, \citenamefont
  {Silvestri},\ and\ \citenamefont {Zhao}}]{Wang:2023tjj}%
  \BibitemOpen
  \bibfield  {author} {\bibinfo {author} {\bibfnamefont {Z.}~\bibnamefont
  {Wang}}, \bibinfo {author} {\bibfnamefont {S.~H.}\ \bibnamefont
  {Mirpoorian}}, \bibinfo {author} {\bibfnamefont {L.}~\bibnamefont
  {Pogosian}}, \bibinfo {author} {\bibfnamefont {A.}~\bibnamefont
  {Silvestri}},\ and\ \bibinfo {author} {\bibfnamefont {G.-B.}\ \bibnamefont
  {Zhao}},\ }\href@noop {} {\  (\bibinfo {year} {2023})},\ \Eprint
  {https://arxiv.org/abs/2305.05667} {arXiv:2305.05667 [astro-ph.CO]}
  \BibitemShut {NoStop}%
\bibitem [{\citenamefont {Nguyen}\ \emph {et~al.}(2023)\citenamefont {Nguyen},
  \citenamefont {Huterer},\ and\ \citenamefont {Wen}}]{Nguyen:2023fip}%
  \BibitemOpen
  \bibfield  {author} {\bibinfo {author} {\bibfnamefont {N.-M.}\ \bibnamefont
  {Nguyen}}, \bibinfo {author} {\bibfnamefont {D.}~\bibnamefont {Huterer}},\
  and\ \bibinfo {author} {\bibfnamefont {Y.}~\bibnamefont {Wen}},\ }\href@noop
  {} {\  (\bibinfo {year} {2023})},\ \Eprint {https://arxiv.org/abs/2302.01331}
  {arXiv:2302.01331 [astro-ph.CO]} \BibitemShut {NoStop}%
\bibitem [{\citenamefont {Wen}\ \emph {et~al.}(2023)\citenamefont {Wen},
  \citenamefont {Nguyen},\ and\ \citenamefont {Huterer}}]{Wen:2023bcj}%
  \BibitemOpen
  \bibfield  {author} {\bibinfo {author} {\bibfnamefont {Y.}~\bibnamefont
  {Wen}}, \bibinfo {author} {\bibfnamefont {N.-M.}\ \bibnamefont {Nguyen}},\
  and\ \bibinfo {author} {\bibfnamefont {D.}~\bibnamefont {Huterer}},\
  }\href@noop {} {\  (\bibinfo {year} {2023})},\ \Eprint
  {https://arxiv.org/abs/2304.07281} {arXiv:2304.07281 [astro-ph.CO]}
  \BibitemShut {NoStop}%
\bibitem [{\citenamefont {Peebles}\ and\ \citenamefont
  {Ratra}(2003)}]{Peebles:2002gy}%
  \BibitemOpen
  \bibfield  {author} {\bibinfo {author} {\bibfnamefont {P.~J.~E.}\
  \bibnamefont {Peebles}}\ and\ \bibinfo {author} {\bibfnamefont
  {B.}~\bibnamefont {Ratra}},\ }\href
  {https://doi.org/10.1103/RevModPhys.75.559} {\bibfield  {journal} {\bibinfo
  {journal} {Rev. Mod. Phys.}\ }\textbf {\bibinfo {volume} {75}},\ \bibinfo
  {pages} {559} (\bibinfo {year} {2003})},\ \Eprint
  {https://arxiv.org/abs/astro-ph/0207347} {arXiv:astro-ph/0207347}
  \BibitemShut {NoStop}%
\bibitem [{\citenamefont {Mukhanov}(2005)}]{Mukhanov:991646}%
  \BibitemOpen
  \bibfield  {author} {\bibinfo {author} {\bibfnamefont {V.}~\bibnamefont
  {Mukhanov}},\ }\href {https://doi.org/10.1017/CBO9780511790553} {\emph
  {\bibinfo {title} {{Physical Foundations of Cosmology}}}}\ (\bibinfo
  {publisher} {Cambridge Univ. Press},\ \bibinfo {address} {Cambridge},\
  \bibinfo {year} {2005})\BibitemShut {NoStop}%
\bibitem [{\citenamefont {Carr}\ \emph {et~al.}(2016)\citenamefont {Carr},
  \citenamefont {Kuhnel},\ and\ \citenamefont {Sandstad}}]{Carr:2016drx}%
  \BibitemOpen
  \bibfield  {author} {\bibinfo {author} {\bibfnamefont {B.}~\bibnamefont
  {Carr}}, \bibinfo {author} {\bibfnamefont {F.}~\bibnamefont {Kuhnel}},\ and\
  \bibinfo {author} {\bibfnamefont {M.}~\bibnamefont {Sandstad}},\ }\href
  {https://doi.org/10.1103/PhysRevD.94.083504} {\bibfield  {journal} {\bibinfo
  {journal} {Phys. Rev. D}\ }\textbf {\bibinfo {volume} {94}},\ \bibinfo
  {pages} {083504} (\bibinfo {year} {2016})},\ \Eprint
  {https://arxiv.org/abs/1607.06077} {arXiv:1607.06077 [astro-ph.CO]}
  \BibitemShut {NoStop}%
\bibitem [{\citenamefont {Akerib}\ \emph {et~al.}(2017)\citenamefont {Akerib}
  \emph {et~al.}}]{LUX:2016ggv}%
  \BibitemOpen
  \bibfield  {author} {\bibinfo {author} {\bibfnamefont {D.~S.}\ \bibnamefont
  {Akerib}} \emph {et~al.} (\bibinfo {collaboration} {LUX}),\ }\href
  {https://doi.org/10.1103/PhysRevLett.118.021303} {\bibfield  {journal}
  {\bibinfo  {journal} {Phys. Rev. Lett.}\ }\textbf {\bibinfo {volume} {118}},\
  \bibinfo {pages} {021303} (\bibinfo {year} {2017})},\ \Eprint
  {https://arxiv.org/abs/1608.07648} {arXiv:1608.07648 [astro-ph.CO]}
  \BibitemShut {NoStop}%
\bibitem [{\citenamefont {Gaitskell}(2004)}]{Gaitskell:2004gd}%
  \BibitemOpen
  \bibfield  {author} {\bibinfo {author} {\bibfnamefont {R.~J.}\ \bibnamefont
  {Gaitskell}},\ }\href {https://doi.org/10.1146/annurev.nucl.54.070103.181244}
  {\bibfield  {journal} {\bibinfo  {journal} {Ann. Rev. Nucl. Part. Sci.}\
  }\textbf {\bibinfo {volume} {54}},\ \bibinfo {pages} {315} (\bibinfo {year}
  {2004})}\BibitemShut {NoStop}%
\bibitem [{\citenamefont {Riess}\ \emph {et~al.}(1998)\citenamefont {Riess}
  \emph {et~al.}}]{Riess:1998cb}%
  \BibitemOpen
  \bibfield  {author} {\bibinfo {author} {\bibfnamefont {A.~G.}\ \bibnamefont
  {Riess}} \emph {et~al.} (\bibinfo {collaboration} {Supernova Search Team}),\
  }\href {https://doi.org/10.1086/300499} {\bibfield  {journal} {\bibinfo
  {journal} {Astron. J.}\ }\textbf {\bibinfo {volume} {116}},\ \bibinfo {pages}
  {1009} (\bibinfo {year} {1998})},\ \Eprint
  {https://arxiv.org/abs/astro-ph/9805201} {arXiv:astro-ph/9805201}
  \BibitemShut {NoStop}%
\bibitem [{\citenamefont {Perlmutter}\ \emph {et~al.}(1999)\citenamefont
  {Perlmutter} \emph {et~al.}}]{Perlmutter:1998np}%
  \BibitemOpen
  \bibfield  {author} {\bibinfo {author} {\bibfnamefont {S.}~\bibnamefont
  {Perlmutter}} \emph {et~al.} (\bibinfo {collaboration} {Supernova Cosmology
  Project}),\ }\href {https://doi.org/10.1086/307221} {\bibfield  {journal}
  {\bibinfo  {journal} {Astrophys. J.}\ }\textbf {\bibinfo {volume} {517}},\
  \bibinfo {pages} {565} (\bibinfo {year} {1999})},\ \Eprint
  {https://arxiv.org/abs/astro-ph/9812133} {arXiv:astro-ph/9812133}
  \BibitemShut {NoStop}%
\bibitem [{\citenamefont {Guth}(1981)}]{Guth:1980zm}%
  \BibitemOpen
  \bibfield  {author} {\bibinfo {author} {\bibfnamefont {A.~H.}\ \bibnamefont
  {Guth}},\ }\href {https://doi.org/10.1103/PhysRevD.23.347} {\bibfield
  {journal} {\bibinfo  {journal} {Phys. Rev. D}\ }\textbf {\bibinfo {volume}
  {23}},\ \bibinfo {pages} {347} (\bibinfo {year} {1981})}\BibitemShut
  {NoStop}%
\bibitem [{\citenamefont {Linde}(1982)}]{Linde:1981mu}%
  \BibitemOpen
  \bibfield  {author} {\bibinfo {author} {\bibfnamefont {A.~D.}\ \bibnamefont
  {Linde}},\ }\href {https://doi.org/10.1016/0370-2693(82)91219-9} {\bibfield
  {journal} {\bibinfo  {journal} {Phys. Lett. B}\ }\textbf {\bibinfo {volume}
  {108}},\ \bibinfo {pages} {389} (\bibinfo {year} {1982})}\BibitemShut
  {NoStop}%
\bibitem [{\citenamefont {Weinberg}(1989)}]{Weinberg:1988cp}%
  \BibitemOpen
  \bibfield  {author} {\bibinfo {author} {\bibfnamefont {S.}~\bibnamefont
  {Weinberg}},\ }\href {https://doi.org/10.1103/RevModPhys.61.1} {\bibfield
  {journal} {\bibinfo  {journal} {Rev. Mod. Phys.}\ }\textbf {\bibinfo {volume}
  {61}},\ \bibinfo {pages} {1} (\bibinfo {year} {1989})}\BibitemShut {NoStop}%
\bibitem [{\citenamefont {Copeland}\ \emph {et~al.}(2006)\citenamefont
  {Copeland}, \citenamefont {Sami},\ and\ \citenamefont
  {Tsujikawa}}]{Copeland:2006wr}%
  \BibitemOpen
  \bibfield  {author} {\bibinfo {author} {\bibfnamefont {E.~J.}\ \bibnamefont
  {Copeland}}, \bibinfo {author} {\bibfnamefont {M.}~\bibnamefont {Sami}},\
  and\ \bibinfo {author} {\bibfnamefont {S.}~\bibnamefont {Tsujikawa}},\ }\href
  {https://doi.org/10.1142/S021827180600942X} {\bibfield  {journal} {\bibinfo
  {journal} {Int. J. Mod. Phys. D}\ }\textbf {\bibinfo {volume} {15}},\
  \bibinfo {pages} {1753} (\bibinfo {year} {2006})},\ \Eprint
  {https://arxiv.org/abs/hep-th/0603057} {arXiv:hep-th/0603057} \BibitemShut
  {NoStop}%
\bibitem [{\citenamefont {Addazi}\ \emph {et~al.}(2022)\citenamefont {Addazi}
  \emph {et~al.}}]{Addazi:2021xuf}%
  \BibitemOpen
  \bibfield  {author} {\bibinfo {author} {\bibfnamefont {A.}~\bibnamefont
  {Addazi}} \emph {et~al.},\ }\href
  {https://doi.org/10.1016/j.ppnp.2022.103948} {\bibfield  {journal} {\bibinfo
  {journal} {Prog. Part. Nucl. Phys.}\ }\textbf {\bibinfo {volume} {125}},\
  \bibinfo {pages} {103948} (\bibinfo {year} {2022})},\ \Eprint
  {https://arxiv.org/abs/2111.05659} {arXiv:2111.05659 [hep-ph]} \BibitemShut
  {NoStop}%
\bibitem [{\citenamefont {Abdalla}\ \emph {et~al.}(2022)\citenamefont {Abdalla}
  \emph {et~al.}}]{Abdalla:2022yfr}%
  \BibitemOpen
  \bibfield  {author} {\bibinfo {author} {\bibfnamefont {E.}~\bibnamefont
  {Abdalla}} \emph {et~al.},\ }\href
  {https://doi.org/10.1016/j.jheap.2022.04.002} {\bibfield  {journal} {\bibinfo
   {journal} {JHEAp}\ }\textbf {\bibinfo {volume} {34}},\ \bibinfo {pages} {49}
  (\bibinfo {year} {2022})},\ \Eprint {https://arxiv.org/abs/2203.06142}
  {arXiv:2203.06142 [astro-ph.CO]} \BibitemShut {NoStop}%
\bibitem [{\citenamefont {Di~Valentino}\ \emph
  {et~al.}(2021{\natexlab{a}})\citenamefont {Di~Valentino} \emph
  {et~al.}}]{DiValentino:2020vhf}%
  \BibitemOpen
  \bibfield  {author} {\bibinfo {author} {\bibfnamefont {E.}~\bibnamefont
  {Di~Valentino}} \emph {et~al.},\ }\href
  {https://doi.org/10.1016/j.astropartphys.2021.102606} {\bibfield  {journal}
  {\bibinfo  {journal} {Astropart. Phys.}\ }\textbf {\bibinfo {volume} {131}},\
  \bibinfo {pages} {102606} (\bibinfo {year} {2021}{\natexlab{a}})},\ \Eprint
  {https://arxiv.org/abs/2008.11283} {arXiv:2008.11283 [astro-ph.CO]}
  \BibitemShut {NoStop}%
\bibitem [{\citenamefont {Di~Valentino}\ \emph
  {et~al.}(2021{\natexlab{b}})\citenamefont {Di~Valentino} \emph
  {et~al.}}]{DiValentino:2020vvd}%
  \BibitemOpen
  \bibfield  {author} {\bibinfo {author} {\bibfnamefont {E.}~\bibnamefont
  {Di~Valentino}} \emph {et~al.},\ }\href
  {https://doi.org/10.1016/j.astropartphys.2021.102604} {\bibfield  {journal}
  {\bibinfo  {journal} {Astropart. Phys.}\ }\textbf {\bibinfo {volume} {131}},\
  \bibinfo {pages} {102604} (\bibinfo {year} {2021}{\natexlab{b}})},\ \Eprint
  {https://arxiv.org/abs/2008.11285} {arXiv:2008.11285 [astro-ph.CO]}
  \BibitemShut {NoStop}%
\bibitem [{\citenamefont {Staicova}(2021)}]{Staicova:2021ajb}%
  \BibitemOpen
  \bibfield  {author} {\bibinfo {author} {\bibfnamefont {D.}~\bibnamefont
  {Staicova}},\ }in\ \href {https://doi.org/10.1142/9789811269776_0151} {\emph
  {\bibinfo {booktitle} {{16th Marcel Grossmann Meeting on~Recent Developments
  in Theoretical and Experimental General Relativity, Astrophysics and
  Relativistic Field Theories}}}}\ (\bibinfo {year} {2021})\ \Eprint
  {https://arxiv.org/abs/2111.07907} {arXiv:2111.07907 [astro-ph.CO]}
  \BibitemShut {NoStop}%
\bibitem [{\citenamefont {Di~Valentino}\ \emph
  {et~al.}(2021{\natexlab{c}})\citenamefont {Di~Valentino}, \citenamefont
  {Mena}, \citenamefont {Pan}, \citenamefont {Visinelli}, \citenamefont {Yang},
  \citenamefont {Melchiorri}, \citenamefont {Mota}, \citenamefont {Riess},\
  and\ \citenamefont {Silk}}]{DiValentino:2021izs}%
  \BibitemOpen
  \bibfield  {author} {\bibinfo {author} {\bibfnamefont {E.}~\bibnamefont
  {Di~Valentino}}, \bibinfo {author} {\bibfnamefont {O.}~\bibnamefont {Mena}},
  \bibinfo {author} {\bibfnamefont {S.}~\bibnamefont {Pan}}, \bibinfo {author}
  {\bibfnamefont {L.}~\bibnamefont {Visinelli}}, \bibinfo {author}
  {\bibfnamefont {W.}~\bibnamefont {Yang}}, \bibinfo {author} {\bibfnamefont
  {A.}~\bibnamefont {Melchiorri}}, \bibinfo {author} {\bibfnamefont {D.~F.}\
  \bibnamefont {Mota}}, \bibinfo {author} {\bibfnamefont {A.~G.}\ \bibnamefont
  {Riess}},\ and\ \bibinfo {author} {\bibfnamefont {J.}~\bibnamefont {Silk}},\
  }\href {https://doi.org/10.1088/1361-6382/ac086d} {\bibfield  {journal}
  {\bibinfo  {journal} {Class. Quant. Grav.}\ }\textbf {\bibinfo {volume}
  {38}},\ \bibinfo {pages} {153001} (\bibinfo {year} {2021}{\natexlab{c}})},\
  \Eprint {https://arxiv.org/abs/2103.01183} {arXiv:2103.01183 [astro-ph.CO]}
  \BibitemShut {NoStop}%
\bibitem [{\citenamefont {Perivolaropoulos}\ and\ \citenamefont
  {Skara}(2022)}]{Perivolaropoulos:2021jda}%
  \BibitemOpen
  \bibfield  {author} {\bibinfo {author} {\bibfnamefont {L.}~\bibnamefont
  {Perivolaropoulos}}\ and\ \bibinfo {author} {\bibfnamefont {F.}~\bibnamefont
  {Skara}},\ }\href {https://doi.org/10.1016/j.newar.2022.101659} {\bibfield
  {journal} {\bibinfo  {journal} {New Astron. Rev.}\ }\textbf {\bibinfo
  {volume} {95}},\ \bibinfo {pages} {101659} (\bibinfo {year} {2022})},\
  \Eprint {https://arxiv.org/abs/2105.05208} {arXiv:2105.05208 [astro-ph.CO]}
  \BibitemShut {NoStop}%
\bibitem [{\citenamefont {Di~Valentino}\ \emph {et~al.}(2022)\citenamefont
  {Di~Valentino}, \citenamefont {Giar\`e}, \citenamefont {Melchiorri},\ and\
  \citenamefont {Silk}}]{DiValentino:2022oon}%
  \BibitemOpen
  \bibfield  {author} {\bibinfo {author} {\bibfnamefont {E.}~\bibnamefont
  {Di~Valentino}}, \bibinfo {author} {\bibfnamefont {W.}~\bibnamefont
  {Giar\`e}}, \bibinfo {author} {\bibfnamefont {A.}~\bibnamefont
  {Melchiorri}},\ and\ \bibinfo {author} {\bibfnamefont {J.}~\bibnamefont
  {Silk}},\ }\href {https://doi.org/10.1103/PhysRevD.106.103506} {\bibfield
  {journal} {\bibinfo  {journal} {Phys. Rev. D}\ }\textbf {\bibinfo {volume}
  {106}},\ \bibinfo {pages} {103506} (\bibinfo {year} {2022})},\ \Eprint
  {https://arxiv.org/abs/2209.12872} {arXiv:2209.12872 [astro-ph.CO]}
  \BibitemShut {NoStop}%
\bibitem [{\citenamefont {Di~Valentino}\ \emph
  {et~al.}(2021{\natexlab{d}})\citenamefont {Di~Valentino} \emph
  {et~al.}}]{DiValentino:2020zio}%
  \BibitemOpen
  \bibfield  {author} {\bibinfo {author} {\bibfnamefont {E.}~\bibnamefont
  {Di~Valentino}} \emph {et~al.},\ }\href
  {https://doi.org/10.1016/j.astropartphys.2021.102605} {\bibfield  {journal}
  {\bibinfo  {journal} {Astropart. Phys.}\ }\textbf {\bibinfo {volume} {131}},\
  \bibinfo {pages} {102605} (\bibinfo {year} {2021}{\natexlab{d}})},\ \Eprint
  {https://arxiv.org/abs/2008.11284} {arXiv:2008.11284 [astro-ph.CO]}
  \BibitemShut {NoStop}%
\bibitem [{\citenamefont {Sajjad~Athar}\ \emph {et~al.}(2022)\citenamefont
  {Sajjad~Athar} \emph {et~al.}}]{SajjadAthar:2021prg}%
  \BibitemOpen
  \bibfield  {author} {\bibinfo {author} {\bibfnamefont {M.}~\bibnamefont
  {Sajjad~Athar}} \emph {et~al.},\ }\href
  {https://doi.org/10.1016/j.ppnp.2022.103947} {\bibfield  {journal} {\bibinfo
  {journal} {Prog. Part. Nucl. Phys.}\ }\textbf {\bibinfo {volume} {124}},\
  \bibinfo {pages} {103947} (\bibinfo {year} {2022})},\ \Eprint
  {https://arxiv.org/abs/2111.07586} {arXiv:2111.07586 [hep-ph]} \BibitemShut
  {NoStop}%
\bibitem [{\citenamefont {Nunes}\ and\ \citenamefont
  {Vagnozzi}(2021)}]{Nunes:2021ipq}%
  \BibitemOpen
  \bibfield  {author} {\bibinfo {author} {\bibfnamefont {R.~C.}\ \bibnamefont
  {Nunes}}\ and\ \bibinfo {author} {\bibfnamefont {S.}~\bibnamefont
  {Vagnozzi}},\ }\href {https://doi.org/10.1093/mnras/stab1613} {\bibfield
  {journal} {\bibinfo  {journal} {Mon. Not. Roy. Astron. Soc.}\ }\textbf
  {\bibinfo {volume} {505}},\ \bibinfo {pages} {5427} (\bibinfo {year}
  {2021})},\ \Eprint {https://arxiv.org/abs/2106.01208} {arXiv:2106.01208
  [astro-ph.CO]} \BibitemShut {NoStop}%
\bibitem [{\citenamefont {Verde}\ \emph {et~al.}(2019)\citenamefont {Verde},
  \citenamefont {Treu},\ and\ \citenamefont {Riess}}]{Verde:2019ivm}%
  \BibitemOpen
  \bibfield  {author} {\bibinfo {author} {\bibfnamefont {L.}~\bibnamefont
  {Verde}}, \bibinfo {author} {\bibfnamefont {T.}~\bibnamefont {Treu}},\ and\
  \bibinfo {author} {\bibfnamefont {A.~G.}\ \bibnamefont {Riess}},\ }\href
  {https://doi.org/10.1038/s41550-019-0902-0} {\bibfield  {journal} {\bibinfo
  {journal} {Nature Astron.}\ }\textbf {\bibinfo {volume} {3}},\ \bibinfo
  {pages} {891} (\bibinfo {year} {2019})},\ \Eprint
  {https://arxiv.org/abs/1907.10625} {arXiv:1907.10625 [astro-ph.CO]}
  \BibitemShut {NoStop}%
\bibitem [{\citenamefont {Di~Valentino}(2021)}]{DiValentino:2020vnx}%
  \BibitemOpen
  \bibfield  {author} {\bibinfo {author} {\bibfnamefont {E.}~\bibnamefont
  {Di~Valentino}},\ }\href {https://doi.org/10.1093/mnras/stab187} {\bibfield
  {journal} {\bibinfo  {journal} {Mon. Not. Roy. Astron. Soc.}\ }\textbf
  {\bibinfo {volume} {502}},\ \bibinfo {pages} {2065} (\bibinfo {year}
  {2021})},\ \Eprint {https://arxiv.org/abs/2011.00246} {arXiv:2011.00246
  [astro-ph.CO]} \BibitemShut {NoStop}%
\bibitem [{\citenamefont {Riess}(2019)}]{Riess:2019qba}%
  \BibitemOpen
  \bibfield  {author} {\bibinfo {author} {\bibfnamefont {A.~G.}\ \bibnamefont
  {Riess}},\ }\href {https://doi.org/10.1038/s42254-019-0137-0} {\bibfield
  {journal} {\bibinfo  {journal} {Nature Rev. Phys.}\ }\textbf {\bibinfo
  {volume} {2}},\ \bibinfo {pages} {10} (\bibinfo {year} {2019})},\ \Eprint
  {https://arxiv.org/abs/2001.03624} {arXiv:2001.03624 [astro-ph.CO]}
  \BibitemShut {NoStop}%
\bibitem [{\citenamefont {Riess}\ \emph
  {et~al.}(2022{\natexlab{a}})\citenamefont {Riess}, \citenamefont {Breuval},
  \citenamefont {Yuan}, \citenamefont {Casertano}, \citenamefont {Macri},
  \citenamefont {Bowers}, \citenamefont {Scolnic}, \citenamefont
  {Cantat-Gaudin}, \citenamefont {Anderson},\ and\ \citenamefont
  {Reyes}}]{Riess:2022mme}%
  \BibitemOpen
  \bibfield  {author} {\bibinfo {author} {\bibfnamefont {A.~G.}\ \bibnamefont
  {Riess}}, \bibinfo {author} {\bibfnamefont {L.}~\bibnamefont {Breuval}},
  \bibinfo {author} {\bibfnamefont {W.}~\bibnamefont {Yuan}}, \bibinfo {author}
  {\bibfnamefont {S.}~\bibnamefont {Casertano}}, \bibinfo {author}
  {\bibfnamefont {L.~M.}\ \bibnamefont {Macri}}, \bibinfo {author}
  {\bibfnamefont {J.~B.}\ \bibnamefont {Bowers}}, \bibinfo {author}
  {\bibfnamefont {D.}~\bibnamefont {Scolnic}}, \bibinfo {author} {\bibfnamefont
  {T.}~\bibnamefont {Cantat-Gaudin}}, \bibinfo {author} {\bibfnamefont {R.~I.}\
  \bibnamefont {Anderson}},\ and\ \bibinfo {author} {\bibfnamefont {M.~C.}\
  \bibnamefont {Reyes}},\ }\href {https://doi.org/10.3847/1538-4357/ac8f24}
  {\bibfield  {journal} {\bibinfo  {journal} {Astrophys. J.}\ }\textbf
  {\bibinfo {volume} {938}},\ \bibinfo {pages} {36} (\bibinfo {year}
  {2022}{\natexlab{a}})},\ \Eprint {https://arxiv.org/abs/2208.01045}
  {arXiv:2208.01045 [astro-ph.CO]} \BibitemShut {NoStop}%
\bibitem [{\citenamefont {Aghanim}\ \emph
  {et~al.}(2020{\natexlab{a}})\citenamefont {Aghanim} \emph
  {et~al.}}]{Planck:2018vyg}%
  \BibitemOpen
  \bibfield  {author} {\bibinfo {author} {\bibfnamefont {N.}~\bibnamefont
  {Aghanim}} \emph {et~al.} (\bibinfo {collaboration} {Planck}),\ }\href
  {https://doi.org/10.1051/0004-6361/201833910} {\bibfield  {journal} {\bibinfo
   {journal} {Astron. Astrophys.}\ }\textbf {\bibinfo {volume} {641}},\
  \bibinfo {pages} {A6} (\bibinfo {year} {2020}{\natexlab{a}})},\ \bibinfo
  {note} {[Erratum: Astron.Astrophys. 652, C4 (2021)]},\ \Eprint
  {https://arxiv.org/abs/1807.06209} {arXiv:1807.06209 [astro-ph.CO]}
  \BibitemShut {NoStop}%
\bibitem [{\citenamefont {Aiola}\ \emph {et~al.}(2020)\citenamefont {Aiola}
  \emph {et~al.}}]{ACT:2020gnv}%
  \BibitemOpen
  \bibfield  {author} {\bibinfo {author} {\bibfnamefont {S.}~\bibnamefont
  {Aiola}} \emph {et~al.} (\bibinfo {collaboration} {ACT}),\ }\href
  {https://doi.org/10.1088/1475-7516/2020/12/047} {\bibfield  {journal}
  {\bibinfo  {journal} {JCAP}\ }\textbf {\bibinfo {volume} {12}},\ \bibinfo
  {pages} {047}},\ \Eprint {https://arxiv.org/abs/2007.07288} {arXiv:2007.07288
  [astro-ph.CO]} \BibitemShut {NoStop}%
\bibitem [{\citenamefont {Riess}\ \emph
  {et~al.}(2022{\natexlab{b}})\citenamefont {Riess} \emph
  {et~al.}}]{Riess:2021jrx}%
  \BibitemOpen
  \bibfield  {author} {\bibinfo {author} {\bibfnamefont {A.~G.}\ \bibnamefont
  {Riess}} \emph {et~al.},\ }\href {https://doi.org/10.3847/2041-8213/ac5c5b}
  {\bibfield  {journal} {\bibinfo  {journal} {Astrophys. J. Lett.}\ }\textbf
  {\bibinfo {volume} {934}},\ \bibinfo {pages} {L7} (\bibinfo {year}
  {2022}{\natexlab{b}})},\ \Eprint {https://arxiv.org/abs/2112.04510}
  {arXiv:2112.04510 [astro-ph.CO]} \BibitemShut {NoStop}%
\bibitem [{\citenamefont {Wong}\ \emph {et~al.}(2020)\citenamefont {Wong} \emph
  {et~al.}}]{Wong:2019kwg}%
  \BibitemOpen
  \bibfield  {author} {\bibinfo {author} {\bibfnamefont {K.~C.}\ \bibnamefont
  {Wong}} \emph {et~al.},\ }\href {https://doi.org/10.1093/mnras/stz3094}
  {\bibfield  {journal} {\bibinfo  {journal} {Mon. Not. Roy. Astron. Soc.}\
  }\textbf {\bibinfo {volume} {498}},\ \bibinfo {pages} {1420} (\bibinfo {year}
  {2020})},\ \Eprint {https://arxiv.org/abs/1907.04869} {arXiv:1907.04869
  [astro-ph.CO]} \BibitemShut {NoStop}%
\bibitem [{\citenamefont {Huang}\ \emph {et~al.}(2019)\citenamefont {Huang},
  \citenamefont {Riess}, \citenamefont {Yuan}, \citenamefont {Macri},
  \citenamefont {Zakamska}, \citenamefont {Casertano}, \citenamefont
  {Whitelock}, \citenamefont {Hoffmann}, \citenamefont {Filippenko},\ and\
  \citenamefont {Scolnic}}]{Huang:2019yhh}%
  \BibitemOpen
  \bibfield  {author} {\bibinfo {author} {\bibfnamefont {C.~D.}\ \bibnamefont
  {Huang}}, \bibinfo {author} {\bibfnamefont {A.~G.}\ \bibnamefont {Riess}},
  \bibinfo {author} {\bibfnamefont {W.}~\bibnamefont {Yuan}}, \bibinfo {author}
  {\bibfnamefont {L.~M.}\ \bibnamefont {Macri}}, \bibinfo {author}
  {\bibfnamefont {N.~L.}\ \bibnamefont {Zakamska}}, \bibinfo {author}
  {\bibfnamefont {S.}~\bibnamefont {Casertano}}, \bibinfo {author}
  {\bibfnamefont {P.~A.}\ \bibnamefont {Whitelock}}, \bibinfo {author}
  {\bibfnamefont {S.~L.}\ \bibnamefont {Hoffmann}}, \bibinfo {author}
  {\bibfnamefont {A.~V.}\ \bibnamefont {Filippenko}},\ and\ \bibinfo {author}
  {\bibfnamefont {D.}~\bibnamefont {Scolnic}}\ }\href
  {https://doi.org/10.3847/1538-4357/ab5dbd} {10.3847/1538-4357/ab5dbd}
  (\bibinfo {year} {2019}),\ \Eprint {https://arxiv.org/abs/1908.10883}
  {arXiv:1908.10883 [astro-ph.CO]} \BibitemShut {NoStop}%
\bibitem [{\citenamefont {Pesce}\ \emph {et~al.}(2020)\citenamefont {Pesce}
  \emph {et~al.}}]{Pesce:2020xfe}%
  \BibitemOpen
  \bibfield  {author} {\bibinfo {author} {\bibfnamefont {D.~W.}\ \bibnamefont
  {Pesce}} \emph {et~al.},\ }\href {https://doi.org/10.3847/2041-8213/ab75f0}
  {\bibfield  {journal} {\bibinfo  {journal} {Astrophys. J. Lett.}\ }\textbf
  {\bibinfo {volume} {891}},\ \bibinfo {pages} {L1} (\bibinfo {year} {2020})},\
  \Eprint {https://arxiv.org/abs/2001.09213} {arXiv:2001.09213 [astro-ph.CO]}
  \BibitemShut {NoStop}%
\bibitem [{\citenamefont {Kourkchi}\ \emph {et~al.}(2020)\citenamefont
  {Kourkchi}, \citenamefont {Tully}, \citenamefont {Anand}, \citenamefont
  {Courtois}, \citenamefont {Dupuy}, \citenamefont {Neill}, \citenamefont
  {Rizzi},\ and\ \citenamefont {Seibert}}]{Kourkchi:2020iyz}%
  \BibitemOpen
  \bibfield  {author} {\bibinfo {author} {\bibfnamefont {E.}~\bibnamefont
  {Kourkchi}}, \bibinfo {author} {\bibfnamefont {R.~B.}\ \bibnamefont {Tully}},
  \bibinfo {author} {\bibfnamefont {G.~S.}\ \bibnamefont {Anand}}, \bibinfo
  {author} {\bibfnamefont {H.~M.}\ \bibnamefont {Courtois}}, \bibinfo {author}
  {\bibfnamefont {A.}~\bibnamefont {Dupuy}}, \bibinfo {author} {\bibfnamefont
  {J.~D.}\ \bibnamefont {Neill}}, \bibinfo {author} {\bibfnamefont
  {L.}~\bibnamefont {Rizzi}},\ and\ \bibinfo {author} {\bibfnamefont
  {M.}~\bibnamefont {Seibert}},\ }\href
  {https://doi.org/10.3847/1538-4357/ab901c} {\bibfield  {journal} {\bibinfo
  {journal} {Astrophys. J.}\ }\textbf {\bibinfo {volume} {896}},\ \bibinfo
  {pages} {3} (\bibinfo {year} {2020})},\ \Eprint
  {https://arxiv.org/abs/2004.14499} {arXiv:2004.14499 [astro-ph.GA]}
  \BibitemShut {NoStop}%
\bibitem [{\citenamefont {Schombert}\ \emph {et~al.}(2020)\citenamefont
  {Schombert}, \citenamefont {McGaugh},\ and\ \citenamefont
  {Lelli}}]{Schombert:2020pxm}%
  \BibitemOpen
  \bibfield  {author} {\bibinfo {author} {\bibfnamefont {J.}~\bibnamefont
  {Schombert}}, \bibinfo {author} {\bibfnamefont {S.}~\bibnamefont {McGaugh}},\
  and\ \bibinfo {author} {\bibfnamefont {F.}~\bibnamefont {Lelli}},\ }\href
  {https://doi.org/10.3847/1538-3881/ab9d88} {\bibfield  {journal} {\bibinfo
  {journal} {Astron. J.}\ }\textbf {\bibinfo {volume} {160}},\ \bibinfo {pages}
  {71} (\bibinfo {year} {2020})},\ \Eprint {https://arxiv.org/abs/2006.08615}
  {arXiv:2006.08615 [astro-ph.CO]} \BibitemShut {NoStop}%
\bibitem [{\citenamefont {Blakeslee}\ \emph {et~al.}(2021)\citenamefont
  {Blakeslee}, \citenamefont {Jensen}, \citenamefont {Ma}, \citenamefont
  {Milne},\ and\ \citenamefont {Greene}}]{Blakeslee:2021rqi}%
  \BibitemOpen
  \bibfield  {author} {\bibinfo {author} {\bibfnamefont {J.~P.}\ \bibnamefont
  {Blakeslee}}, \bibinfo {author} {\bibfnamefont {J.~B.}\ \bibnamefont
  {Jensen}}, \bibinfo {author} {\bibfnamefont {C.-P.}\ \bibnamefont {Ma}},
  \bibinfo {author} {\bibfnamefont {P.~A.}\ \bibnamefont {Milne}},\ and\
  \bibinfo {author} {\bibfnamefont {J.~E.}\ \bibnamefont {Greene}},\ }\href
  {https://doi.org/10.3847/1538-4357/abe86a} {\bibfield  {journal} {\bibinfo
  {journal} {Astrophys. J.}\ }\textbf {\bibinfo {volume} {911}},\ \bibinfo
  {pages} {65} (\bibinfo {year} {2021})},\ \Eprint
  {https://arxiv.org/abs/2101.02221} {arXiv:2101.02221 [astro-ph.CO]}
  \BibitemShut {NoStop}%
\bibitem [{\citenamefont {de~Jaeger}\ \emph {et~al.}(2022)\citenamefont
  {de~Jaeger}, \citenamefont {Galbany}, \citenamefont {Riess}, \citenamefont
  {Stahl}, \citenamefont {Shappee}, \citenamefont {Filippenko},\ and\
  \citenamefont {Zheng}}]{deJaeger:2022lit}%
  \BibitemOpen
  \bibfield  {author} {\bibinfo {author} {\bibfnamefont {T.}~\bibnamefont
  {de~Jaeger}}, \bibinfo {author} {\bibfnamefont {L.}~\bibnamefont {Galbany}},
  \bibinfo {author} {\bibfnamefont {A.~G.}\ \bibnamefont {Riess}}, \bibinfo
  {author} {\bibfnamefont {B.~E.}\ \bibnamefont {Stahl}}, \bibinfo {author}
  {\bibfnamefont {B.~J.}\ \bibnamefont {Shappee}}, \bibinfo {author}
  {\bibfnamefont {A.~V.}\ \bibnamefont {Filippenko}},\ and\ \bibinfo {author}
  {\bibfnamefont {W.}~\bibnamefont {Zheng}},\ }\href
  {https://doi.org/10.1093/mnras/stac1661} {\bibfield  {journal} {\bibinfo
  {journal} {Mon. Not. Roy. Astron. Soc.}\ }\textbf {\bibinfo {volume} {514}},\
  \bibinfo {pages} {4620} (\bibinfo {year} {2022})},\ \Eprint
  {https://arxiv.org/abs/2203.08974} {arXiv:2203.08974 [astro-ph.CO]}
  \BibitemShut {NoStop}%
\bibitem [{\citenamefont {Shajib}\ \emph {et~al.}(2023)\citenamefont {Shajib}
  \emph {et~al.}}]{Shajib:2023uig}%
  \BibitemOpen
  \bibfield  {author} {\bibinfo {author} {\bibfnamefont {A.~J.}\ \bibnamefont
  {Shajib}} \emph {et~al.},\ }\href
  {https://doi.org/10.1051/0004-6361/202345878} {\bibfield  {journal} {\bibinfo
   {journal} {Astron. Astrophys.}\ }\textbf {\bibinfo {volume} {673}},\
  \bibinfo {pages} {A9} (\bibinfo {year} {2023})},\ \Eprint
  {https://arxiv.org/abs/2301.02656} {arXiv:2301.02656 [astro-ph.CO]}
  \BibitemShut {NoStop}%
\bibitem [{\citenamefont {Scolnic}\ \emph {et~al.}(2023)\citenamefont
  {Scolnic}, \citenamefont {Riess}, \citenamefont {Wu}, \citenamefont {Li},
  \citenamefont {Anand}, \citenamefont {Beaton}, \citenamefont {Casertano},
  \citenamefont {Anderson}, \citenamefont {Dhawan},\ and\ \citenamefont
  {Ke}}]{Scolnic:2023mrv}%
  \BibitemOpen
  \bibfield  {author} {\bibinfo {author} {\bibfnamefont {D.}~\bibnamefont
  {Scolnic}}, \bibinfo {author} {\bibfnamefont {A.~G.}\ \bibnamefont {Riess}},
  \bibinfo {author} {\bibfnamefont {J.}~\bibnamefont {Wu}}, \bibinfo {author}
  {\bibfnamefont {S.}~\bibnamefont {Li}}, \bibinfo {author} {\bibfnamefont
  {G.~S.}\ \bibnamefont {Anand}}, \bibinfo {author} {\bibfnamefont
  {R.}~\bibnamefont {Beaton}}, \bibinfo {author} {\bibfnamefont
  {S.}~\bibnamefont {Casertano}}, \bibinfo {author} {\bibfnamefont
  {R.}~\bibnamefont {Anderson}}, \bibinfo {author} {\bibfnamefont
  {S.}~\bibnamefont {Dhawan}},\ and\ \bibinfo {author} {\bibfnamefont
  {X.}~\bibnamefont {Ke}},\ }\href@noop {} {\  (\bibinfo {year} {2023})},\
  \Eprint {https://arxiv.org/abs/2304.06693} {arXiv:2304.06693 [astro-ph.CO]}
  \BibitemShut {NoStop}%
\bibitem [{\citenamefont {Anderson}\ \emph {et~al.}(2023)\citenamefont
  {Anderson}, \citenamefont {Koblischke},\ and\ \citenamefont
  {Eyer}}]{Anderson:2023aga}%
  \BibitemOpen
  \bibfield  {author} {\bibinfo {author} {\bibfnamefont {R.~I.}\ \bibnamefont
  {Anderson}}, \bibinfo {author} {\bibfnamefont {N.~W.}\ \bibnamefont
  {Koblischke}},\ and\ \bibinfo {author} {\bibfnamefont {L.}~\bibnamefont
  {Eyer}},\ }\href@noop {} {\  (\bibinfo {year} {2023})},\ \Eprint
  {https://arxiv.org/abs/2303.04790} {arXiv:2303.04790 [astro-ph.CO]}
  \BibitemShut {NoStop}%
\bibitem [{\citenamefont {Freedman}(2021)}]{Freedman:2021ahq}%
  \BibitemOpen
  \bibfield  {author} {\bibinfo {author} {\bibfnamefont {W.~L.}\ \bibnamefont
  {Freedman}},\ }\href {https://doi.org/10.3847/1538-4357/ac0e95} {\bibfield
  {journal} {\bibinfo  {journal} {Astrophys. J.}\ }\textbf {\bibinfo {volume}
  {919}},\ \bibinfo {pages} {16} (\bibinfo {year} {2021})},\ \Eprint
  {https://arxiv.org/abs/2106.15656} {arXiv:2106.15656 [astro-ph.CO]}
  \BibitemShut {NoStop}%
\bibitem [{\citenamefont {Freedman}\ \emph {et~al.}(2020)\citenamefont
  {Freedman}, \citenamefont {Madore}, \citenamefont {Hoyt}, \citenamefont
  {Jang}, \citenamefont {Beaton}, \citenamefont {Lee}, \citenamefont {Monson},
  \citenamefont {Neeley},\ and\ \citenamefont {Rich}}]{Freedman:2020dne}%
  \BibitemOpen
  \bibfield  {author} {\bibinfo {author} {\bibfnamefont {W.~L.}\ \bibnamefont
  {Freedman}}, \bibinfo {author} {\bibfnamefont {B.~F.}\ \bibnamefont
  {Madore}}, \bibinfo {author} {\bibfnamefont {T.}~\bibnamefont {Hoyt}},
  \bibinfo {author} {\bibfnamefont {I.~S.}\ \bibnamefont {Jang}}, \bibinfo
  {author} {\bibfnamefont {R.}~\bibnamefont {Beaton}}, \bibinfo {author}
  {\bibfnamefont {M.~G.}\ \bibnamefont {Lee}}, \bibinfo {author} {\bibfnamefont
  {A.}~\bibnamefont {Monson}}, \bibinfo {author} {\bibfnamefont
  {J.}~\bibnamefont {Neeley}},\ and\ \bibinfo {author} {\bibfnamefont
  {J.}~\bibnamefont {Rich}}\ }\href {https://doi.org/10.3847/1538-4357/ab7339}
  {10.3847/1538-4357/ab7339} (\bibinfo {year} {2020}),\ \Eprint
  {https://arxiv.org/abs/2002.01550} {arXiv:2002.01550 [astro-ph.GA]}
  \BibitemShut {NoStop}%
\bibitem [{\citenamefont {Heymans}\ \emph {et~al.}(2021)\citenamefont {Heymans}
  \emph {et~al.}}]{Heymans:2020gsg}%
  \BibitemOpen
  \bibfield  {author} {\bibinfo {author} {\bibfnamefont {C.}~\bibnamefont
  {Heymans}} \emph {et~al.},\ }\href
  {https://doi.org/10.1051/0004-6361/202039063} {\bibfield  {journal} {\bibinfo
   {journal} {Astron. Astrophys.}\ }\textbf {\bibinfo {volume} {646}},\
  \bibinfo {pages} {A140} (\bibinfo {year} {2021})},\ \Eprint
  {https://arxiv.org/abs/2007.15632} {arXiv:2007.15632 [astro-ph.CO]}
  \BibitemShut {NoStop}%
\bibitem [{\citenamefont {Abbott}\ \emph {et~al.}(2022)\citenamefont {Abbott}
  \emph {et~al.}}]{DES:2021wwk}%
  \BibitemOpen
  \bibfield  {author} {\bibinfo {author} {\bibfnamefont {T.~M.~C.}\
  \bibnamefont {Abbott}} \emph {et~al.} (\bibinfo {collaboration} {DES}),\
  }\href {https://doi.org/10.1103/PhysRevD.105.023520} {\bibfield  {journal}
  {\bibinfo  {journal} {Phys. Rev. D}\ }\textbf {\bibinfo {volume} {105}},\
  \bibinfo {pages} {023520} (\bibinfo {year} {2022})},\ \Eprint
  {https://arxiv.org/abs/2105.13549} {arXiv:2105.13549 [astro-ph.CO]}
  \BibitemShut {NoStop}%
\bibitem [{\citenamefont {Dalal}\ \emph {et~al.}(2023)\citenamefont {Dalal}
  \emph {et~al.}}]{Dalal:2023olq}%
  \BibitemOpen
  \bibfield  {author} {\bibinfo {author} {\bibfnamefont {R.}~\bibnamefont
  {Dalal}} \emph {et~al.},\ }\href@noop {} {\  (\bibinfo {year} {2023})},\
  \Eprint {https://arxiv.org/abs/2304.00701} {arXiv:2304.00701 [astro-ph.CO]}
  \BibitemShut {NoStop}%
\bibitem [{\citenamefont {Balkenhol}\ \emph {et~al.}(2022)\citenamefont
  {Balkenhol} \emph {et~al.}}]{SPT-3G:2022hvq}%
  \BibitemOpen
  \bibfield  {author} {\bibinfo {author} {\bibfnamefont {L.}~\bibnamefont
  {Balkenhol}} \emph {et~al.} (\bibinfo {collaboration} {SPT-3G}),\ }\href@noop
  {} {\  (\bibinfo {year} {2022})},\ \Eprint {https://arxiv.org/abs/2212.05642}
  {arXiv:2212.05642 [astro-ph.CO]} \BibitemShut {NoStop}%
\bibitem [{\citenamefont {Brout}\ \emph {et~al.}(2022)\citenamefont {Brout}
  \emph {et~al.}}]{Brout:2021mpj}%
  \BibitemOpen
  \bibfield  {author} {\bibinfo {author} {\bibfnamefont {D.}~\bibnamefont
  {Brout}} \emph {et~al.},\ }\href {https://doi.org/10.3847/1538-4357/ac8bcc}
  {\bibfield  {journal} {\bibinfo  {journal} {Astrophys. J.}\ }\textbf
  {\bibinfo {volume} {938}},\ \bibinfo {pages} {111} (\bibinfo {year}
  {2022})},\ \Eprint {https://arxiv.org/abs/2112.03864} {arXiv:2112.03864
  [astro-ph.CO]} \BibitemShut {NoStop}%
\bibitem [{\citenamefont {Knox}\ and\ \citenamefont
  {Millea}(2020)}]{Knox:2019rjx}%
  \BibitemOpen
  \bibfield  {author} {\bibinfo {author} {\bibfnamefont {L.}~\bibnamefont
  {Knox}}\ and\ \bibinfo {author} {\bibfnamefont {M.}~\bibnamefont {Millea}},\
  }\href {https://doi.org/10.1103/PhysRevD.101.043533} {\bibfield  {journal}
  {\bibinfo  {journal} {Phys. Rev. D}\ }\textbf {\bibinfo {volume} {101}},\
  \bibinfo {pages} {043533} (\bibinfo {year} {2020})},\ \Eprint
  {https://arxiv.org/abs/1908.03663} {arXiv:1908.03663 [astro-ph.CO]}
  \BibitemShut {NoStop}%
\bibitem [{\citenamefont {Jedamzik}\ \emph {et~al.}(2021)\citenamefont
  {Jedamzik}, \citenamefont {Pogosian},\ and\ \citenamefont
  {Zhao}}]{Jedamzik:2020zmd}%
  \BibitemOpen
  \bibfield  {author} {\bibinfo {author} {\bibfnamefont {K.}~\bibnamefont
  {Jedamzik}}, \bibinfo {author} {\bibfnamefont {L.}~\bibnamefont {Pogosian}},\
  and\ \bibinfo {author} {\bibfnamefont {G.-B.}\ \bibnamefont {Zhao}},\ }\href
  {https://doi.org/10.1038/s42005-021-00628-x} {\bibfield  {journal} {\bibinfo
  {journal} {Commun. in Phys.}\ }\textbf {\bibinfo {volume} {4}},\ \bibinfo
  {pages} {123} (\bibinfo {year} {2021})},\ \Eprint
  {https://arxiv.org/abs/2010.04158} {arXiv:2010.04158 [astro-ph.CO]}
  \BibitemShut {NoStop}%
\bibitem [{\citenamefont {Poulin}\ \emph {et~al.}(2023)\citenamefont {Poulin},
  \citenamefont {Smith},\ and\ \citenamefont {Karwal}}]{Poulin:2023lkg}%
  \BibitemOpen
  \bibfield  {author} {\bibinfo {author} {\bibfnamefont {V.}~\bibnamefont
  {Poulin}}, \bibinfo {author} {\bibfnamefont {T.~L.}\ \bibnamefont {Smith}},\
  and\ \bibinfo {author} {\bibfnamefont {T.}~\bibnamefont {Karwal}},\
  }\href@noop {} {\  (\bibinfo {year} {2023})},\ \Eprint
  {https://arxiv.org/abs/2302.09032} {arXiv:2302.09032 [astro-ph.CO]}
  \BibitemShut {NoStop}%
\bibitem [{\citenamefont {Di~Valentino}(2022)}]{DiValentino:2022fjm}%
  \BibitemOpen
  \bibfield  {author} {\bibinfo {author} {\bibfnamefont {E.}~\bibnamefont
  {Di~Valentino}},\ }\href {https://doi.org/10.3390/universe8080399} {\bibfield
   {journal} {\bibinfo  {journal} {Universe}\ }\textbf {\bibinfo {volume}
  {8}},\ \bibinfo {pages} {399} (\bibinfo {year} {2022})}\BibitemShut {NoStop}%
\bibitem [{\citenamefont {Krishnan}\ \emph {et~al.}(2021)\citenamefont
  {Krishnan}, \citenamefont {Colg\'ain}, \citenamefont {Sheikh-Jabbari},\ and\
  \citenamefont {Yang}}]{Krishnan:2020vaf}%
  \BibitemOpen
  \bibfield  {author} {\bibinfo {author} {\bibfnamefont {C.}~\bibnamefont
  {Krishnan}}, \bibinfo {author} {\bibfnamefont {E.~O.}\ \bibnamefont
  {Colg\'ain}}, \bibinfo {author} {\bibfnamefont {M.~M.}\ \bibnamefont
  {Sheikh-Jabbari}},\ and\ \bibinfo {author} {\bibfnamefont {T.}~\bibnamefont
  {Yang}},\ }\href {https://doi.org/10.1103/PhysRevD.103.103509} {\bibfield
  {journal} {\bibinfo  {journal} {Phys. Rev. D}\ }\textbf {\bibinfo {volume}
  {103}},\ \bibinfo {pages} {103509} (\bibinfo {year} {2021})},\ \Eprint
  {https://arxiv.org/abs/2011.02858} {arXiv:2011.02858 [astro-ph.CO]}
  \BibitemShut {NoStop}%
\bibitem [{\citenamefont {Adil}\ \emph {et~al.}(2023)\citenamefont {Adil},
  \citenamefont {Akarsu}, \citenamefont {Malekjani}, \citenamefont {Colg\'ain},
  \citenamefont {Pourojaghi}, \citenamefont {Sen},\ and\ \citenamefont
  {Sheikh-Jabbari}}]{Adil:2023jtu}%
  \BibitemOpen
  \bibfield  {author} {\bibinfo {author} {\bibfnamefont {S.~A.}\ \bibnamefont
  {Adil}}, \bibinfo {author} {\bibfnamefont {O.}~\bibnamefont {Akarsu}},
  \bibinfo {author} {\bibfnamefont {M.}~\bibnamefont {Malekjani}}, \bibinfo
  {author} {\bibfnamefont {E.~O.}\ \bibnamefont {Colg\'ain}}, \bibinfo {author}
  {\bibfnamefont {S.}~\bibnamefont {Pourojaghi}}, \bibinfo {author}
  {\bibfnamefont {A.~A.}\ \bibnamefont {Sen}},\ and\ \bibinfo {author}
  {\bibfnamefont {M.~M.}\ \bibnamefont {Sheikh-Jabbari}},\ }\href@noop {} {\
  (\bibinfo {year} {2023})},\ \Eprint {https://arxiv.org/abs/2303.06928}
  {arXiv:2303.06928 [astro-ph.CO]} \BibitemShut {NoStop}%
\bibitem [{\citenamefont {Saridakis}\ \emph {et~al.}(2021)\citenamefont
  {Saridakis} \emph {et~al.}}]{CANTATA:2021ktz}%
  \BibitemOpen
  \bibfield  {author} {\bibinfo {author} {\bibfnamefont {E.~N.}\ \bibnamefont
  {Saridakis}} \emph {et~al.} (\bibinfo {collaboration} {CANTATA}),\
  }\href@noop {} {\  (\bibinfo {year} {2021})},\ \Eprint
  {https://arxiv.org/abs/2105.12582} {arXiv:2105.12582 [gr-qc]} \BibitemShut
  {NoStop}%
\bibitem [{\citenamefont {Kamionkowski}\ and\ \citenamefont
  {Riess}(2022)}]{Kamionkowski:2022pkx}%
  \BibitemOpen
  \bibfield  {author} {\bibinfo {author} {\bibfnamefont {M.}~\bibnamefont
  {Kamionkowski}}\ and\ \bibinfo {author} {\bibfnamefont {A.~G.}\ \bibnamefont
  {Riess}},\ }\href@noop {} {\  (\bibinfo {year} {2022})},\ \Eprint
  {https://arxiv.org/abs/2211.04492} {arXiv:2211.04492 [astro-ph.CO]}
  \BibitemShut {NoStop}%
\bibitem [{\citenamefont {Linder}(2005)}]{Linder:2005in}%
  \BibitemOpen
  \bibfield  {author} {\bibinfo {author} {\bibfnamefont {E.~V.}\ \bibnamefont
  {Linder}},\ }\href {https://doi.org/10.1103/PhysRevD.72.043529} {\bibfield
  {journal} {\bibinfo  {journal} {Phys. Rev. D}\ }\textbf {\bibinfo {volume}
  {72}},\ \bibinfo {pages} {043529} (\bibinfo {year} {2005})},\ \Eprint
  {https://arxiv.org/abs/astro-ph/0507263} {arXiv:astro-ph/0507263}
  \BibitemShut {NoStop}%
\bibitem [{\citenamefont {Linder}\ and\ \citenamefont
  {Jenkins}(2003)}]{Linder:2003dr}%
  \BibitemOpen
  \bibfield  {author} {\bibinfo {author} {\bibfnamefont {E.~V.}\ \bibnamefont
  {Linder}}\ and\ \bibinfo {author} {\bibfnamefont {A.}~\bibnamefont
  {Jenkins}},\ }\href {https://doi.org/10.1046/j.1365-2966.2003.07112.x}
  {\bibfield  {journal} {\bibinfo  {journal} {Mon. Not. Roy. Astron. Soc.}\
  }\textbf {\bibinfo {volume} {346}},\ \bibinfo {pages} {573} (\bibinfo {year}
  {2003})},\ \Eprint {https://arxiv.org/abs/astro-ph/0305286}
  {arXiv:astro-ph/0305286} \BibitemShut {NoStop}%
\bibitem [{\citenamefont {Linder}(2003{\natexlab{a}})}]{Linder:2002et}%
  \BibitemOpen
  \bibfield  {author} {\bibinfo {author} {\bibfnamefont {E.~V.}\ \bibnamefont
  {Linder}},\ }\href {https://doi.org/10.1103/PhysRevLett.90.091301} {\bibfield
   {journal} {\bibinfo  {journal} {Phys. Rev. Lett.}\ }\textbf {\bibinfo
  {volume} {90}},\ \bibinfo {pages} {091301} (\bibinfo {year}
  {2003}{\natexlab{a}})},\ \Eprint {https://arxiv.org/abs/astro-ph/0208512}
  {arXiv:astro-ph/0208512} \BibitemShut {NoStop}%
\bibitem [{\citenamefont {Wang}\ and\ \citenamefont
  {Steinhardt}(1998)}]{Wang:1998gt}%
  \BibitemOpen
  \bibfield  {author} {\bibinfo {author} {\bibfnamefont {L.-M.}\ \bibnamefont
  {Wang}}\ and\ \bibinfo {author} {\bibfnamefont {P.~J.}\ \bibnamefont
  {Steinhardt}},\ }\href {https://doi.org/10.1086/306436} {\bibfield  {journal}
  {\bibinfo  {journal} {Astrophys. J.}\ }\textbf {\bibinfo {volume} {508}},\
  \bibinfo {pages} {483} (\bibinfo {year} {1998})},\ \Eprint
  {https://arxiv.org/abs/astro-ph/9804015} {arXiv:astro-ph/9804015}
  \BibitemShut {NoStop}%
\bibitem [{\citenamefont {Linder}(2004)}]{Linder:2004ng}%
  \BibitemOpen
  \bibfield  {author} {\bibinfo {author} {\bibfnamefont {E.~V.}\ \bibnamefont
  {Linder}},\ }\href {https://doi.org/10.1103/PhysRevD.70.023511} {\bibfield
  {journal} {\bibinfo  {journal} {Phys. Rev. D}\ }\textbf {\bibinfo {volume}
  {70}},\ \bibinfo {pages} {023511} (\bibinfo {year} {2004})},\ \Eprint
  {https://arxiv.org/abs/astro-ph/0402503} {arXiv:astro-ph/0402503}
  \BibitemShut {NoStop}%
\bibitem [{\citenamefont {Linder}\ and\ \citenamefont
  {Polarski}(2019)}]{Linder:2018pth}%
  \BibitemOpen
  \bibfield  {author} {\bibinfo {author} {\bibfnamefont {E.~V.}\ \bibnamefont
  {Linder}}\ and\ \bibinfo {author} {\bibfnamefont {D.}~\bibnamefont
  {Polarski}},\ }\href {https://doi.org/10.1103/PhysRevD.99.023503} {\bibfield
  {journal} {\bibinfo  {journal} {Phys. Rev. D}\ }\textbf {\bibinfo {volume}
  {99}},\ \bibinfo {pages} {023503} (\bibinfo {year} {2019})},\ \Eprint
  {https://arxiv.org/abs/1810.10547} {arXiv:1810.10547 [astro-ph.CO]}
  \BibitemShut {NoStop}%
\bibitem [{\citenamefont {Linder}\ and\ \citenamefont
  {Cahn}(2007)}]{Linder:2007hg}%
  \BibitemOpen
  \bibfield  {author} {\bibinfo {author} {\bibfnamefont {E.~V.}\ \bibnamefont
  {Linder}}\ and\ \bibinfo {author} {\bibfnamefont {R.~N.}\ \bibnamefont
  {Cahn}},\ }\href {https://doi.org/10.1016/j.astropartphys.2007.09.003}
  {\bibfield  {journal} {\bibinfo  {journal} {Astropart. Phys.}\ }\textbf
  {\bibinfo {volume} {28}},\ \bibinfo {pages} {481} (\bibinfo {year} {2007})},\
  \Eprint {https://arxiv.org/abs/astro-ph/0701317} {arXiv:astro-ph/0701317}
  \BibitemShut {NoStop}%
\bibitem [{\citenamefont {Linder}(2003{\natexlab{b}})}]{Linder:2003ze}%
  \BibitemOpen
  \bibfield  {author} {\bibinfo {author} {\bibfnamefont {E.~V.}\ \bibnamefont
  {Linder}},\ }\href {https://doi.org/10.1063/1.1543500} {\bibfield  {journal}
  {\bibinfo  {journal} {AIP Conf. Proc.}\ }\textbf {\bibinfo {volume} {655}},\
  \bibinfo {pages} {193} (\bibinfo {year} {2003}{\natexlab{b}})},\ \Eprint
  {https://arxiv.org/abs/astro-ph/0302038} {arXiv:astro-ph/0302038}
  \BibitemShut {NoStop}%
\bibitem [{\citenamefont {Linder}(2023)}]{Linder:2023klx}%
  \BibitemOpen
  \bibfield  {author} {\bibinfo {author} {\bibfnamefont {E.~V.}\ \bibnamefont
  {Linder}},\ }\href@noop {} {\  (\bibinfo {year} {2023})},\ \Eprint
  {https://arxiv.org/abs/2304.04803} {arXiv:2304.04803 [astro-ph.CO]}
  \BibitemShut {NoStop}%
\bibitem [{\citenamefont {Linder}(2006)}]{Linder:2006xb}%
  \BibitemOpen
  \bibfield  {author} {\bibinfo {author} {\bibfnamefont {E.~V.}\ \bibnamefont
  {Linder}},\ }\href {https://doi.org/10.1016/j.astropartphys.2006.05.004}
  {\bibfield  {journal} {\bibinfo  {journal} {Astropart. Phys.}\ }\textbf
  {\bibinfo {volume} {26}},\ \bibinfo {pages} {102} (\bibinfo {year} {2006})},\
  \Eprint {https://arxiv.org/abs/astro-ph/0604280} {arXiv:astro-ph/0604280}
  \BibitemShut {NoStop}%
\bibitem [{\citenamefont {Lewis}\ \emph {et~al.}(2000)\citenamefont {Lewis},
  \citenamefont {Challinor},\ and\ \citenamefont {Lasenby}}]{Lewis:1999bs}%
  \BibitemOpen
  \bibfield  {author} {\bibinfo {author} {\bibfnamefont {A.}~\bibnamefont
  {Lewis}}, \bibinfo {author} {\bibfnamefont {A.}~\bibnamefont {Challinor}},\
  and\ \bibinfo {author} {\bibfnamefont {A.}~\bibnamefont {Lasenby}},\ }\href
  {https://doi.org/10.1086/309179} {\bibfield  {journal} {\bibinfo  {journal}
  {Astrophys. J.}\ }\textbf {\bibinfo {volume} {538}},\ \bibinfo {pages} {473}
  (\bibinfo {year} {2000})},\ \Eprint {https://arxiv.org/abs/astro-ph/9911177}
  {arXiv:astro-ph/9911177} \BibitemShut {NoStop}%
\bibitem [{\citenamefont {Howlett}\ \emph {et~al.}(2012)\citenamefont
  {Howlett}, \citenamefont {Lewis}, \citenamefont {Hall},\ and\ \citenamefont
  {Challinor}}]{Howlett:2012mh}%
  \BibitemOpen
  \bibfield  {author} {\bibinfo {author} {\bibfnamefont {C.}~\bibnamefont
  {Howlett}}, \bibinfo {author} {\bibfnamefont {A.}~\bibnamefont {Lewis}},
  \bibinfo {author} {\bibfnamefont {A.}~\bibnamefont {Hall}},\ and\ \bibinfo
  {author} {\bibfnamefont {A.}~\bibnamefont {Challinor}},\ }\href
  {https://doi.org/10.1088/1475-7516/2012/04/027} {\bibfield  {journal}
  {\bibinfo  {journal} {JCAP}\ }\textbf {\bibinfo {volume} {04}},\ \bibinfo
  {pages} {027}},\ \Eprint {https://arxiv.org/abs/1201.3654} {arXiv:1201.3654
  [astro-ph.CO]} \BibitemShut {NoStop}%
\bibitem [{\citenamefont {Xu}(2013)}]{Xu:2013tsa}%
  \BibitemOpen
  \bibfield  {author} {\bibinfo {author} {\bibfnamefont {L.}~\bibnamefont
  {Xu}},\ }\href {https://doi.org/10.1103/PhysRevD.88.084032} {\bibfield
  {journal} {\bibinfo  {journal} {Phys. Rev. D}\ }\textbf {\bibinfo {volume}
  {88}},\ \bibinfo {pages} {084032} (\bibinfo {year} {2013})},\ \Eprint
  {https://arxiv.org/abs/1306.2683} {arXiv:1306.2683 [astro-ph.CO]}
  \BibitemShut {NoStop}%
\bibitem [{\citenamefont {Sakr}(2023)}]{Sakr:2023bms}%
  \BibitemOpen
  \bibfield  {author} {\bibinfo {author} {\bibfnamefont {Z.}~\bibnamefont
  {Sakr}}\ }(\bibinfo {year} {2023})\ \Eprint
  {https://arxiv.org/abs/2305.02863} {arXiv:2305.02863 [astro-ph.CO]}
  \BibitemShut {NoStop}%
\bibitem [{gam()}]{gamma_prime_growth}%
  \BibitemOpen
  \href@noop {} {}\bibinfo {howpublished}
  {\url{https://github.com/MinhMPA/CAMB_GammaPrime_Growth}}\BibitemShut
  {NoStop}%
\bibitem [{\citenamefont {Hanson}\ \emph {et~al.}(2010)\citenamefont {Hanson},
  \citenamefont {Challinor},\ and\ \citenamefont {Lewis}}]{Hanson:2009kr}%
  \BibitemOpen
  \bibfield  {author} {\bibinfo {author} {\bibfnamefont {D.}~\bibnamefont
  {Hanson}}, \bibinfo {author} {\bibfnamefont {A.}~\bibnamefont {Challinor}},\
  and\ \bibinfo {author} {\bibfnamefont {A.}~\bibnamefont {Lewis}},\ }\href
  {https://doi.org/10.1007/s10714-010-1036-y} {\bibfield  {journal} {\bibinfo
  {journal} {Gen. Rel. Grav.}\ }\textbf {\bibinfo {volume} {42}},\ \bibinfo
  {pages} {2197} (\bibinfo {year} {2010})},\ \Eprint
  {https://arxiv.org/abs/0911.0612} {arXiv:0911.0612 [astro-ph.CO]}
  \BibitemShut {NoStop}%
\bibitem [{\citenamefont {Lewis}\ and\ \citenamefont
  {Bridle}(2002)}]{Lewis:2002ah}%
  \BibitemOpen
  \bibfield  {author} {\bibinfo {author} {\bibfnamefont {A.}~\bibnamefont
  {Lewis}}\ and\ \bibinfo {author} {\bibfnamefont {S.}~\bibnamefont {Bridle}},\
  }\href {https://doi.org/10.1103/PhysRevD.66.103511} {\bibfield  {journal}
  {\bibinfo  {journal} {Phys. Rev. D}\ }\textbf {\bibinfo {volume} {66}},\
  \bibinfo {pages} {103511} (\bibinfo {year} {2002})},\ \Eprint
  {https://arxiv.org/abs/astro-ph/0205436} {arXiv:astro-ph/0205436}
  \BibitemShut {NoStop}%
\bibitem [{\citenamefont {Lewis}(2013)}]{Lewis:2013hha}%
  \BibitemOpen
  \bibfield  {author} {\bibinfo {author} {\bibfnamefont {A.}~\bibnamefont
  {Lewis}},\ }\href {https://doi.org/10.1103/PhysRevD.87.103529} {\bibfield
  {journal} {\bibinfo  {journal} {Phys. Rev. D}\ }\textbf {\bibinfo {volume}
  {87}},\ \bibinfo {pages} {103529} (\bibinfo {year} {2013})},\ \Eprint
  {https://arxiv.org/abs/1304.4473} {arXiv:1304.4473 [astro-ph.CO]}
  \BibitemShut {NoStop}%
\bibitem [{\citenamefont {Neal}(2005)}]{neal2005taking}%
  \BibitemOpen
  \bibfield  {author} {\bibinfo {author} {\bibfnamefont {R.~M.}\ \bibnamefont
  {Neal}},\ }\href@noop {} {\bibinfo {title} {Taking bigger metropolis steps by
  dragging fast variables}} (\bibinfo {year} {2005}),\ \Eprint
  {https://arxiv.org/abs/math/0502099} {arXiv:math/0502099 [math.ST]}
  \BibitemShut {NoStop}%
\bibitem [{\citenamefont {Aghanim}\ \emph
  {et~al.}(2020{\natexlab{b}})\citenamefont {Aghanim} \emph
  {et~al.}}]{Planck:2018nkj}%
  \BibitemOpen
  \bibfield  {author} {\bibinfo {author} {\bibfnamefont {N.}~\bibnamefont
  {Aghanim}} \emph {et~al.} (\bibinfo {collaboration} {Planck}),\ }\href
  {https://doi.org/10.1051/0004-6361/201833880} {\bibfield  {journal} {\bibinfo
   {journal} {Astron. Astrophys.}\ }\textbf {\bibinfo {volume} {641}},\
  \bibinfo {pages} {A1} (\bibinfo {year} {2020}{\natexlab{b}})},\ \Eprint
  {https://arxiv.org/abs/1807.06205} {arXiv:1807.06205 [astro-ph.CO]}
  \BibitemShut {NoStop}%
\bibitem [{\citenamefont {Aghanim}\ \emph
  {et~al.}(2020{\natexlab{c}})\citenamefont {Aghanim} \emph
  {et~al.}}]{Planck:2019nip}%
  \BibitemOpen
  \bibfield  {author} {\bibinfo {author} {\bibfnamefont {N.}~\bibnamefont
  {Aghanim}} \emph {et~al.} (\bibinfo {collaboration} {Planck}),\ }\href
  {https://doi.org/10.1051/0004-6361/201936386} {\bibfield  {journal} {\bibinfo
   {journal} {Astron. Astrophys.}\ }\textbf {\bibinfo {volume} {641}},\
  \bibinfo {pages} {A5} (\bibinfo {year} {2020}{\natexlab{c}})},\ \Eprint
  {https://arxiv.org/abs/1907.12875} {arXiv:1907.12875 [astro-ph.CO]}
  \BibitemShut {NoStop}%
\bibitem [{\citenamefont {Bennett}\ \emph {et~al.}(2013)\citenamefont {Bennett}
  \emph {et~al.}}]{WMAP:2012fli}%
  \BibitemOpen
  \bibfield  {author} {\bibinfo {author} {\bibfnamefont {C.~L.}\ \bibnamefont
  {Bennett}} \emph {et~al.} (\bibinfo {collaboration} {WMAP}),\ }\href
  {https://doi.org/10.1088/0067-0049/208/2/20} {\bibfield  {journal} {\bibinfo
  {journal} {Astrophys. J. Suppl.}\ }\textbf {\bibinfo {volume} {208}},\
  \bibinfo {pages} {20} (\bibinfo {year} {2013})},\ \Eprint
  {https://arxiv.org/abs/1212.5225} {arXiv:1212.5225 [astro-ph.CO]}
  \BibitemShut {NoStop}%
\bibitem [{\citenamefont {Choi}\ \emph {et~al.}(2020)\citenamefont {Choi} \emph
  {et~al.}}]{ACT:2020frw}%
  \BibitemOpen
  \bibfield  {author} {\bibinfo {author} {\bibfnamefont {S.~K.}\ \bibnamefont
  {Choi}} \emph {et~al.} (\bibinfo {collaboration} {ACT}),\ }\href
  {https://doi.org/10.1088/1475-7516/2020/12/045} {\bibfield  {journal}
  {\bibinfo  {journal} {JCAP}\ }\textbf {\bibinfo {volume} {12}},\ \bibinfo
  {pages} {045}},\ \Eprint {https://arxiv.org/abs/2007.07289} {arXiv:2007.07289
  [astro-ph.CO]} \BibitemShut {NoStop}%
\bibitem [{\citenamefont {Dutcher}\ \emph {et~al.}(2021)\citenamefont {Dutcher}
  \emph {et~al.}}]{SPT-3G:2021eoc}%
  \BibitemOpen
  \bibfield  {author} {\bibinfo {author} {\bibfnamefont {D.}~\bibnamefont
  {Dutcher}} \emph {et~al.} (\bibinfo {collaboration} {SPT-3G}),\ }\href
  {https://doi.org/10.1103/PhysRevD.104.022003} {\bibfield  {journal} {\bibinfo
   {journal} {Phys. Rev. D}\ }\textbf {\bibinfo {volume} {104}},\ \bibinfo
  {pages} {022003} (\bibinfo {year} {2021})},\ \Eprint
  {https://arxiv.org/abs/2101.01684} {arXiv:2101.01684 [astro-ph.CO]}
  \BibitemShut {NoStop}%
\bibitem [{\citenamefont {Alam}\ \emph {et~al.}(2021)\citenamefont {Alam} \emph
  {et~al.}}]{eBOSS:2020yzd}%
  \BibitemOpen
  \bibfield  {author} {\bibinfo {author} {\bibfnamefont {S.}~\bibnamefont
  {Alam}} \emph {et~al.} (\bibinfo {collaboration} {eBOSS}),\ }\href
  {https://doi.org/10.1103/PhysRevD.103.083533} {\bibfield  {journal} {\bibinfo
   {journal} {Phys. Rev. D}\ }\textbf {\bibinfo {volume} {103}},\ \bibinfo
  {pages} {083533} (\bibinfo {year} {2021})},\ \Eprint
  {https://arxiv.org/abs/2007.08991} {arXiv:2007.08991 [astro-ph.CO]}
  \BibitemShut {NoStop}%
\bibitem [{\citenamefont {Gelman}\ and\ \citenamefont {Rubin}(1992)}]{gelman}%
  \BibitemOpen
  \bibfield  {author} {\bibinfo {author} {\bibfnamefont {A.}~\bibnamefont
  {Gelman}}\ and\ \bibinfo {author} {\bibfnamefont {D.~B.}\ \bibnamefont
  {Rubin}},\ }\href {https://doi.org/10.1214/ss/1177011136} {\bibfield
  {journal} {\bibinfo  {journal} {Statistical Science}\ }\textbf {\bibinfo
  {volume} {7}},\ \bibinfo {pages} {457 } (\bibinfo {year} {1992})}\BibitemShut
  {NoStop}%
\bibitem [{\citenamefont {Calabrese}\ \emph {et~al.}(2008)\citenamefont
  {Calabrese}, \citenamefont {Slosar}, \citenamefont {Melchiorri},
  \citenamefont {Smoot},\ and\ \citenamefont {Zahn}}]{Calabrese:2008rt}%
  \BibitemOpen
  \bibfield  {author} {\bibinfo {author} {\bibfnamefont {E.}~\bibnamefont
  {Calabrese}}, \bibinfo {author} {\bibfnamefont {A.}~\bibnamefont {Slosar}},
  \bibinfo {author} {\bibfnamefont {A.}~\bibnamefont {Melchiorri}}, \bibinfo
  {author} {\bibfnamefont {G.~F.}\ \bibnamefont {Smoot}},\ and\ \bibinfo
  {author} {\bibfnamefont {O.}~\bibnamefont {Zahn}},\ }\href
  {https://doi.org/10.1103/PhysRevD.77.123531} {\bibfield  {journal} {\bibinfo
  {journal} {Phys. Rev. D}\ }\textbf {\bibinfo {volume} {77}},\ \bibinfo
  {pages} {123531} (\bibinfo {year} {2008})},\ \Eprint
  {https://arxiv.org/abs/0803.2309} {arXiv:0803.2309 [astro-ph]} \BibitemShut
  {NoStop}%
\bibitem [{\citenamefont {Di~Valentino}\ \emph
  {et~al.}(2021{\natexlab{e}})\citenamefont {Di~Valentino}, \citenamefont
  {Melchiorri},\ and\ \citenamefont {Silk}}]{DiValentino:2020hov}%
  \BibitemOpen
  \bibfield  {author} {\bibinfo {author} {\bibfnamefont {E.}~\bibnamefont
  {Di~Valentino}}, \bibinfo {author} {\bibfnamefont {A.}~\bibnamefont
  {Melchiorri}},\ and\ \bibinfo {author} {\bibfnamefont {J.}~\bibnamefont
  {Silk}},\ }\href {https://doi.org/10.3847/2041-8213/abe1c4} {\bibfield
  {journal} {\bibinfo  {journal} {Astrophys. J. Lett.}\ }\textbf {\bibinfo
  {volume} {908}},\ \bibinfo {pages} {L9} (\bibinfo {year}
  {2021}{\natexlab{e}})},\ \Eprint {https://arxiv.org/abs/2003.04935}
  {arXiv:2003.04935 [astro-ph.CO]} \BibitemShut {NoStop}%
\bibitem [{\citenamefont {Di~Valentino}\ \emph {et~al.}(2019)\citenamefont
  {Di~Valentino}, \citenamefont {Melchiorri},\ and\ \citenamefont
  {Silk}}]{DiValentino:2019qzk}%
  \BibitemOpen
  \bibfield  {author} {\bibinfo {author} {\bibfnamefont {E.}~\bibnamefont
  {Di~Valentino}}, \bibinfo {author} {\bibfnamefont {A.}~\bibnamefont
  {Melchiorri}},\ and\ \bibinfo {author} {\bibfnamefont {J.}~\bibnamefont
  {Silk}},\ }\href {https://doi.org/10.1038/s41550-019-0906-9} {\bibfield
  {journal} {\bibinfo  {journal} {Nature Astron.}\ }\textbf {\bibinfo {volume}
  {4}},\ \bibinfo {pages} {196} (\bibinfo {year} {2019})},\ \Eprint
  {https://arxiv.org/abs/1911.02087} {arXiv:1911.02087 [astro-ph.CO]}
  \BibitemShut {NoStop}%
\bibitem [{\citenamefont {Handley}(2021)}]{Handley:2019tkm}%
  \BibitemOpen
  \bibfield  {author} {\bibinfo {author} {\bibfnamefont {W.}~\bibnamefont
  {Handley}},\ }\href {https://doi.org/10.1103/PhysRevD.103.L041301} {\bibfield
   {journal} {\bibinfo  {journal} {Phys. Rev. D}\ }\textbf {\bibinfo {volume}
  {103}},\ \bibinfo {pages} {L041301} (\bibinfo {year} {2021})},\ \Eprint
  {https://arxiv.org/abs/1908.09139} {arXiv:1908.09139 [astro-ph.CO]}
  \BibitemShut {NoStop}%
\bibitem [{\citenamefont {Aghanim}\ \emph
  {et~al.}(2020{\natexlab{d}})\citenamefont {Aghanim} \emph
  {et~al.}}]{Planck:2018lbu}%
  \BibitemOpen
  \bibfield  {author} {\bibinfo {author} {\bibfnamefont {N.}~\bibnamefont
  {Aghanim}} \emph {et~al.} (\bibinfo {collaboration} {Planck}),\ }\href
  {https://doi.org/10.1051/0004-6361/201833886} {\bibfield  {journal} {\bibinfo
   {journal} {Astron. Astrophys.}\ }\textbf {\bibinfo {volume} {641}},\
  \bibinfo {pages} {A8} (\bibinfo {year} {2020}{\natexlab{d}})},\ \Eprint
  {https://arxiv.org/abs/1807.06210} {arXiv:1807.06210 [astro-ph.CO]}
  \BibitemShut {NoStop}%
\end{thebibliography}%

\newpage
\pagebreak

\appendix

\section{CMB observables in \code{MGCAMB} and \code{CAMB\_GammaPrime\_Growth}}
\label{app:code_comparison}

In this appendix, we provide a direct comparison between the CMB observables as produced by the two codes: \code{MGCAMB} and \code{CAMB\_GammaPrime\_Growth}.
Our aim is to understand the difference between their CMB predictions, which in turn drives the difference in their constraints on $\gamma_L$.
We therefore examine a) the primary, i.e. unlensed, CMB temperature-temperature (TT) angular power spectrum $C^{\mathrm{TT}}_{\ell}$ and b) the CMB lensing potential angular power spectrum $C^{\phi\phi}_{\ell}$. We focus on the Planck(+lensing) dataset(s) and likelihood(s) as our case study. Unless explicitly stated otherwise, we adopt the best-fit base-$\Lambda$CDM cosmology in \cite{Planck:2018vyg}, specifically the ``\code{Plik} best fit'' column in their Table 1, for the following comparison.

Let us first investigate the \emph{unlensed} CMB TT angular power spectrum as an example of primary CMB observables in the two codes.
\begin{figure*}
   \includegraphics[width=0.7\textwidth]{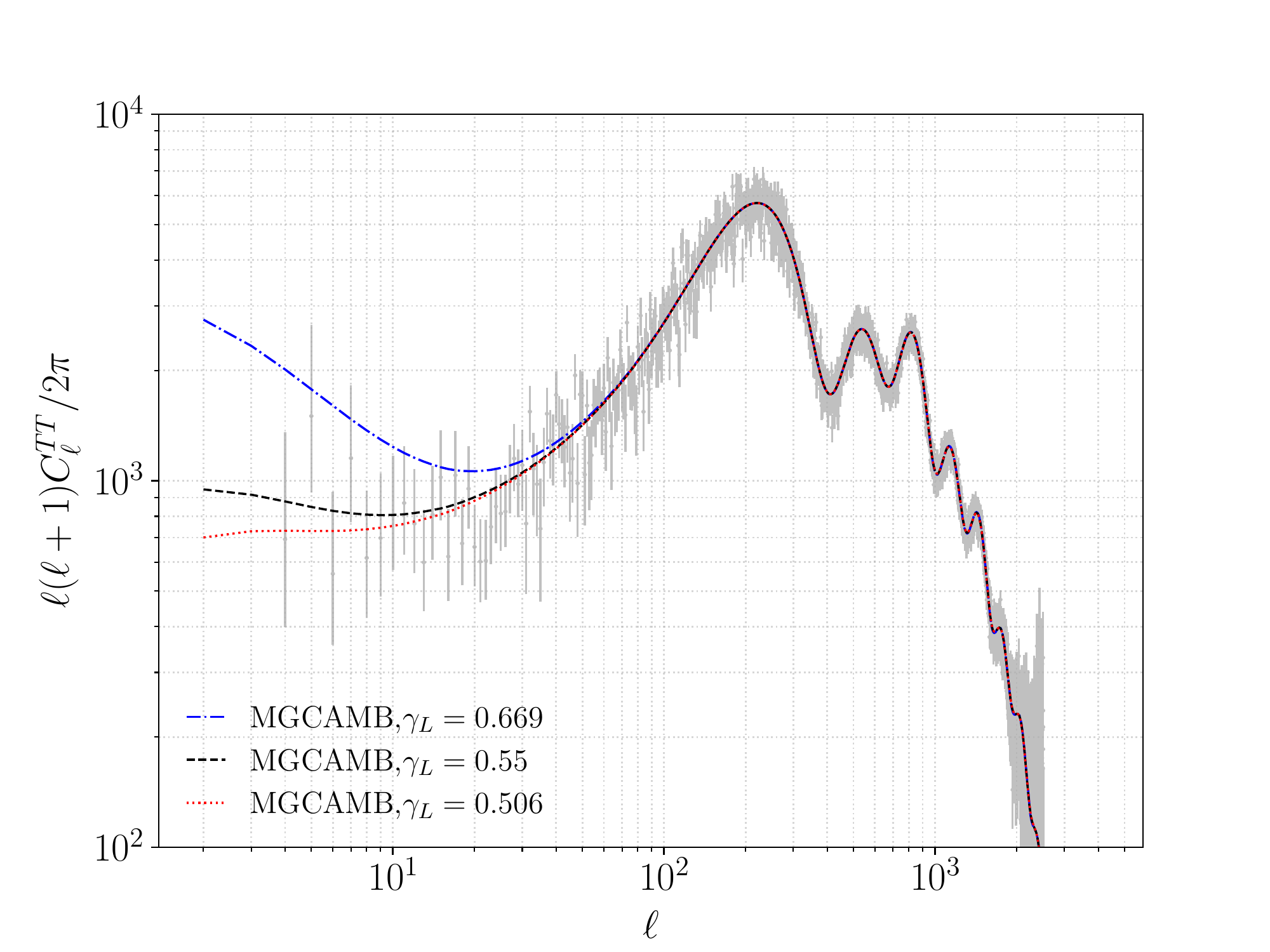}
     \caption{The \emph{unlensed} CMB TT angular power spectrum as predicted by \code{MGCAMB} at the same best-fit Planck 2018 cosmology, but for different values of $\gamma_L$.  Blue and red lines correspond to the cases of $\gamma_L=0.669$ and $\gamma_L=0.506$ -- the mean of $\gamma_L$ constraints reported in the ``Planck+lensing'' columns in \reftab{Pl-MG} and \reftab{Pl-Minh} -- respectively. The black line denotes the GR case of $\gamma_L=0.550$. The Planck 2018 measurements and their associated uncertainties are shown as grey points with error bars.}
    \label{fig:unlensed_TT_Dell_MGCAMB}
\end{figure*}
As discussed in \refsec{codes}, with \code{MGCAMB}, changes in $\gamma_L$ do imply changes in primary CMB observables. Specifically, the variation is significant in the low$\ell$ regime, as shown in \reffig{unlensed_TT_Dell_MGCAMB}. 
On the other hand, with \code{CAMB\_GammaPrime\_Growth}, no change in $\gamma_L$ would alter the primary CMB observables, including the TT angular power spectrum. This is illustrated in \reffig{unlensed_TT_Dell_CAMB_GammaPrime_Growth}.
\begin{figure*}
   \includegraphics[width=0.7\textwidth]{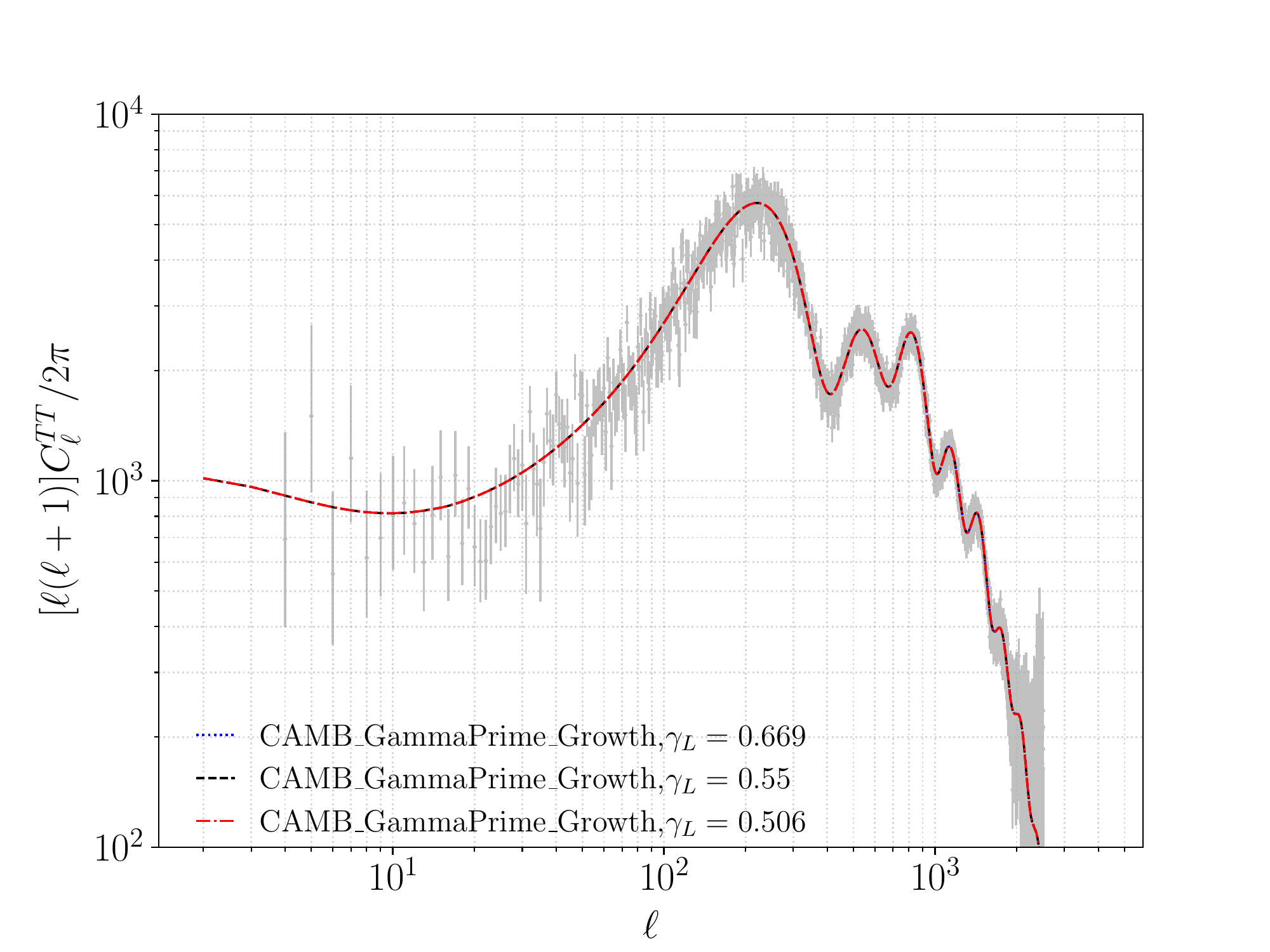}
     \caption{Same as \reffig{unlensed_TT_Dell_MGCAMB}, but using \code{CAMB\_GammaPrime\_Growth}.}
    \label{fig:unlensed_TT_Dell_CAMB_GammaPrime_Growth}
\end{figure*}

\begin{figure*}
   \includegraphics[width=0.7\textwidth]{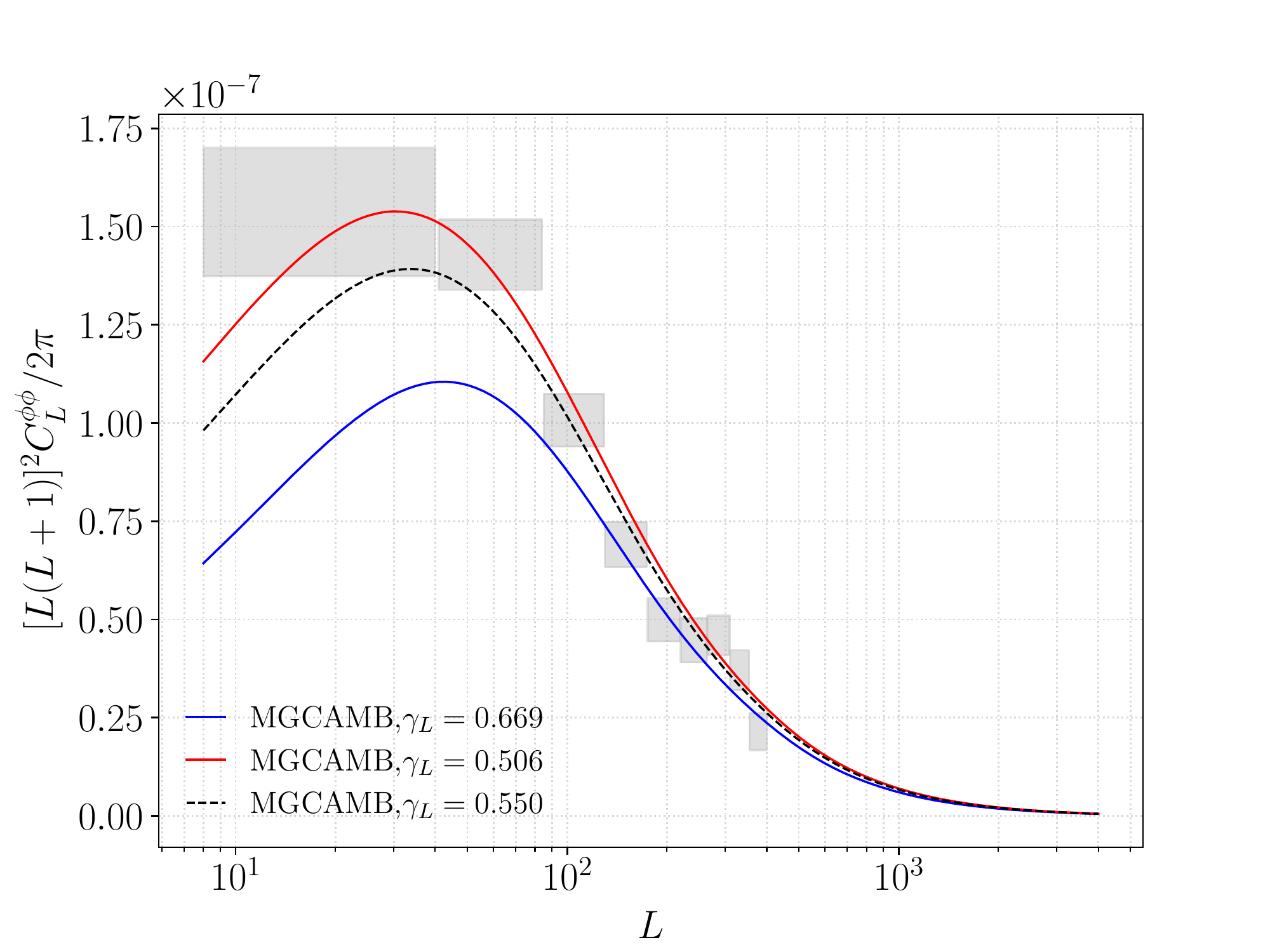}
     \caption{The CMB lensing potential angular power spectrum as predicted by \code{MGCAMB} at the same best-fit Planck 2018 cosmology, for the same values of $\gamma_L$ in \reffigs{unlensed_TT_Dell_MGCAMB}{unlensed_TT_Dell_CAMB_GammaPrime_Growth} (following the same color notation). The Planck 2018 minimum-variance (conservative) estimates of the CMB lensing potential bandpowers and their associated uncertainties are shown as grey boxes.}
    \label{fig:lensing_phiphi_Dell_MGCAMB}
\end{figure*}
\begin{figure*}
   \includegraphics[width=0.7\textwidth]{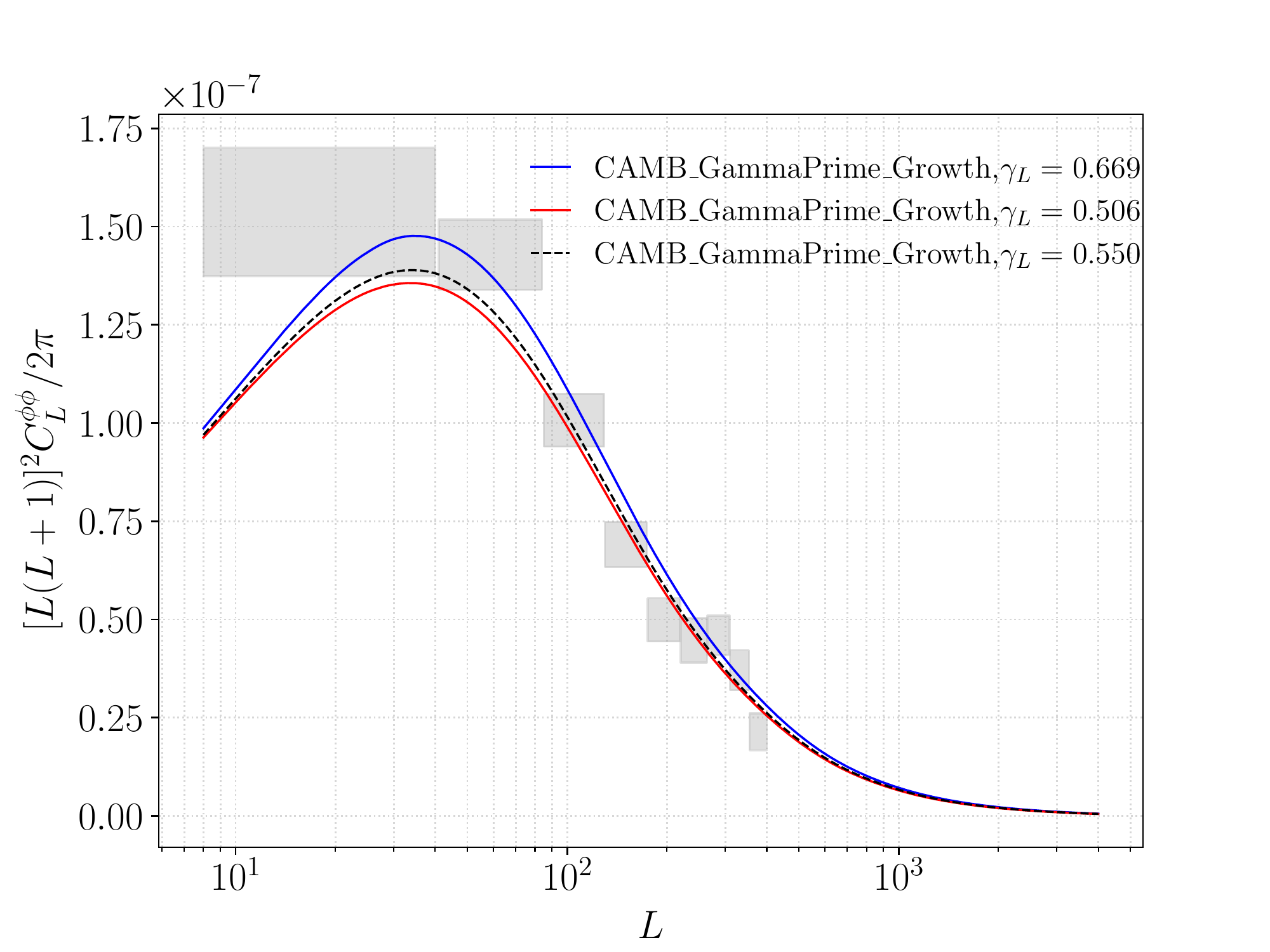}
     \caption{Same as \reffig{unlensed_TT_Dell_MGCAMB}, but using \code{CAMB\_GammaPrime\_Growth}. As explained in \refsec{codes}, with \code{CAMB\_GammaPrime\_Growth}, no change in $\gamma_L$ would alter the primary CMB observables, including the TT angular power spectrum.}
    \label{fig:lensing_phiphi_Dell_CAMB_GammaPrime_Growth}
\end{figure*}
Next, we look at the CMB lensing potentials, which should be sensitive to variations of $\gamma_L$ in both codes.
Comparing \reffigs{lensing_phiphi_Dell_MGCAMB}{lensing_phiphi_Dell_CAMB_GammaPrime_Growth} and the Planck 2018 CMB lensing potential angular power spectrum~\cite{Planck:2018lbu}, it is evident why \code{MGCAMB} prefers $\gamma_L=0.506$ while \code{CAMB\_GammaPrime\_Growth} prefers $\gamma_L=0.669$. Both preferences are driven by the fit to the estimated $C^{\phi\phi}_{\ell}$.
Upon further investigation with \code{MGCAMB} authors\footnote{Private communication.}, we have isolated and attributed the anomalous low-$\ell$ behavior in $C^{\phi\phi}_{\ell}$ from \code{MGCAMB} to a spurious early integrated Sachs-Wolfe (ISW) contribution from a term that contains a time-derivative instance of the modified-gravity parameter $\mu$, i.e. $\dot{\mu}$. As pointed out and discussed in the main text, around \refeq{MGCAMB_gamma_mu}, and in this appendix, this early ISW contribution significantly alters the CMB power spectra on super-horizon scales. That, in turn, shifts the data preference for $\gamma_L$. As this feature was only discovered recently, we caution the community to use \code{MGCAMB} with the $\gamma_L$ parameterization until a fix is released. We note that other modified-gravity parameterizations in \code{MGCAMB}, which in fact are what \code{MGCAMB} was originally intended for and recommended by \code{MGCAMB} authors themselves\footnote{Private communication.}, are still self-consistent and should be employed instead.

Finally, we have verified that adopting either best-fit cosmologies obtained by \code{MGCAMB} or \code{CAMB\_GammaPrime\_Growth} with ``Planck+lensing'' data sets (instead of ``Plik best fit'') does not affect our conclusions from this comparison.

 \end{document}